\documentclass[10pt]{article}
\usepackage{fancyhdr}
\usepackage{extramarks}
\usepackage{amsmath}
\usepackage{amsthm}
\usepackage{amsfonts}
\usepackage{siunitx}
\usepackage{tikz}
\usepackage[plain]{algorithm}
\usepackage{algpseudocode}
\usepackage{multirow}
\usepackage{booktabs}
\usepackage{palatino}
\usepackage{graphicx}
\usepackage{subfigure}
\usepackage[colorlinks,linkcolor=black,anchorcolor=black,citecolor=black,urlcolor=blue]{hyperref}
\usepackage{amsmath,bm}
\usepackage{booktabs}
\usepackage{mathtools}
\usepackage{amssymb}
\usepackage{caption}
\usepackage{capt-of}
\usepackage{mciteplus}
\usepackage{cite}
\usepackage{mathrsfs}
\usepackage[title,titletoc,toc]{appendix}
\usepackage{xr}
\usepackage{parskip}
\usepackage{soul}
\usepackage{textcomp}
\usepackage[colaction]{multicol}
\usepackage[switch]{lineno}
\usepackage{lipsum}
\usepackage{etoolbox}
\usepackage{longtable}
\usepackage{array}
\usepackage{tablefootnote}
\newcolumntype{C}[1]{>{\centering\arraybackslash}p{#1}}
\usetikzlibrary{automata,positioning}
\usetikzlibrary{automata,positioning}

\begin{document}

\title{Vaccine-escape and fast-growing  mutations in the United Kingdom, the United States, Singapore, Spain, South Africa, and other COVID-19-devastated countries}
% \author{Jiahui Chen\orcidA$^1$, Kaifu Gao\orcidB$^1$\footnote{Jiahui Chen and Kaifu Gao contributed equally.}, Rui Wang\orcidC$^1$, Duc Duy Nguyen\orcidD$^2$, and Guo-Wei Wei\orcidE$^{1,3,4}$\footnote{
% 		Corresponding author.		Email: wei@math.msu.edu} \\% Author name
% $^1$ Department of Mathematics, \\
% Michigan State University, MI 48824, USA.\\
% $^2$ Department of Mathematics, \\
% University of Kentucky, KY 40506, USA.\\
% $^3$ Department of Electrical and Computer Engineering,\\
% Michigan State University, MI 48824, USA. \\
% $^4$ Department of Biochemistry and Molecular Biology,\\
% Michigan State University, MI 48824, USA. \\
% }

\author{Rui Wang$^1$, Jiahui Chen$^1$, Kaifu Gao$^1$, and Guo-Wei Wei$^{1,2,3}$\footnote{
		Corresponding author.		Email: weig@msu.edu} \\% Author name
$^1$ Department of Mathematics, \\
Michigan State University, MI 48824, USA.\\
$^2$ Department of Electrical and Computer Engineering,\\
Michigan State University, MI 48824, USA. \\
$^3$ Department of Biochemistry and Molecular Biology,\\
Michigan State University, MI 48824, USA. \\
}
\date{\today} % Date for the report

\maketitle

\begin{abstract}
 Recently, the SARS-CoV-2 variants from the United Kingdom (UK), South Africa, and Brazil have received much attention for their increased infectivity, potentially high virulence, and possible threats to existing vaccines and antibody therapies. The question remains if there are other more infectious variants transmitted around the world. We carry out a large-scale study of 252,874 SARS-CoV-2 genome isolates from patients to identify many other rapidly growing mutations on the spike (S) protein receptor-binding domain (RDB). We reveal that 88 out of 95 significant mutations that were observed more than 10 times strengthen the binding between the RBD and the host angiotensin-converting enzyme 2 (ACE2), indicating the virus evolves toward more infectious variants. In particular, we discover   new fast-growing RBD mutations N439K, L452R, S477N, S477R, and N501T that also enhance the RBD and ACE2 binding. We further unveil that mutation N501Y involved in United Kingdom (UK), South Africa, and Brazil variants may moderately weaken the binding between the RBD and many known antibodies, while mutations E484K and K417N found in South Africa and Brazilian variants can potentially disrupt the binding between the RDB and many known antibodies. Among three newly identified fast-growing RBD mutations,  L452R, { which is now known as part of the California variant B.1.427},   and N501T are able to effectively weaken the binding of many known antibodies with the RBD. 
{Finally, we hypothesize that RBD mutations that can simultaneously make SARS-CoV-2 more infectious and disrupt the existing antibodies, called vaccine escape mutations, will pose an imminent threat to the current crop of vaccines. A list of most likely vaccine escape mutations is given, including N501Y, L452R, E484K, N501T, S494P, and K417N.  
%T478K*, and V367F*. S477N*,  N439K*,
}
Our comprehensive genetic analysis and protein-protein binding study show that the genetic evolution of SARS-CoV-2 on the RBD, which may be regulated by host gene editing, viral proofreading,  random genetic drift, and natural selection, gives rise to more infectious variants that will potentially compromise existing vaccines and antibody therapies.

\end{abstract}

\pagenumbering{roman}
\begin{verbatim}
\end{verbatim}

%{\setcounter{tocdepth}{4} \tableofcontents}
%
  \newpage
 %\clearpage
 %\pagebreak

\setcounter{page}{1}
\renewcommand{\thepage}{{\arabic{page}}}

% \begin{multicols}{2}
% \multicollinenumbers
% \linenumbers
\section{Introduction}
Up to February 19, 2021, coronavirus disease 2019 (COVID-19) caused by severe acute respiratory
syndrome coronavirus 2 (SARS-CoV-2) has taken 2,438,287 lives and infected 109,969,891 people according to the data from World Health Organization (WHO). The first complete SARS-CoV-2 genome sequence was deposited to the GenBank (Access number: NC\_045512.2) on January 5, 2020. Thereafter, new SARS-Cov-2 genome sequences were accumulated rapidly at the GenBank and GISAID, which laid the foundations for analyzing the SARS-CoV-2 mutations, virulence, pathogenicity, antigenicity, and transmissibility. A complete SARS-CoV-2 genome is an unsegmented positive-sense single-stranded RNA virus, which encodes 29 structural and non-structural proteins (NSPs) by its 29,903 nucleotides. NSPs play vital roles in RNA replication, while structure proteins form the viral particle. There are four structural proteins on SARS-CoV-2, namely, spike (S), envelope (E), membrane (M), and nucleocapsid (N) proteins \cite{michel2020characterization,helmy2020covid,naqvi2020insights,mu2020sars}. Among them, the S protein with 1273 residues of SARS-CoV-2 has drawn much attention due to its critic role in viral infection and the development of vaccines and antibody drugs.

The SARS-CoV-2 enters the host cell by interacting between its S protein and the host angiotensin-converting enzyme 2 (ACE2), primed by host transmembrane protease, serine 2 (TMPRSS2) \cite{hoffmann2020sars}. Such a process initiates the response from the host adaptive immune system, which generates antibodies to combat the invading virus. Therefore, the S protein of SARS-CoV-2 has become a target in the development of antibody therapies and vaccines. A major concern is what are the potential impacts of S protein mutations on viral infectivity, the existing vaccines, and antibody therapies.

The most well-known mechanism of mutations is the random genetic drift, which plays a role in the processes of transcription, translation, replication, etc. Compared with DNA viruses, RNA viruses are more prone to random mutations. Unlike other RNA viruses,  such as influenza,  SARS-CoV-2 has a genetic proofreading mechanism regulated by NSP14 and NSP12 (a.k.a RNA-dependent RNA polymerase) \cite{sevajol2014insights,ferron2018structural}, which enables SARS-CoV-2 to have a higher fidelity in its replication. However, the host gene editing was found to be the major source for existing  SARS-CoV-2 mutations \cite{wang2020host}, counting for 65\% of reported mutations. Therefore, the worldwide transmission of COVID-19 provides  SARS-CoV-2 an abundant opportunity to experience fast mutations. Another important mechanism for SARS-CoV-2 evolution is natural selection, which makes the virus more infectious while less virulent, in general \cite{sanjuan2016mechanisms,grubaugh2020making}.  

It has been established that the infectivity of different viral variants in host cells is proportional to the binding free energy (BFE) between the RBD of each variant and the ACE2 \cite{li2005bats, qu2005identification, song2005cross,hoffmann2020sars,walls2020structure}. Based on such a principle, it has been reported that mutations on the S protein have strengthened SARS-CoV-2 infectivity \cite{chen2020mutations}. Whereas, virulent can be a result of mutations on many SARS-CoV-2 proteins. The widely spread asymptomatic COVID-19 infection and transmission can be a result of mutation-induced virulent changes \cite{wang2020decoding}. 
 
Recently, the United Kingdom (UK) variant B.1.1.7 (a.k.a 20I/501Y.V1) \cite{tang2020emergence}, the South Africa variant B.1.351  (a.k.a 20H/501Y.V2) \cite{mwenda2021detection}, and the Brazil(ian) variant P.1 (a.k.a 20J/501Y.V3) \cite{faria2021genomic} have been circulating worldwide, including the United States (US) and Spain. These variants contain mutations on the S protein RBD and are widely speculated to make SARS-CoV-2 more infectious. Specifically, all three variants involve RBD mutation N501Y, whereas the South Africa and Brazil(ian) variants also contain RBD mutations E484K and K417N. 

 An important question is how these new variants will affect the vaccines and antibody drugs. Ideally, this question should be answered by experiments. However, SARS-CoV-2 has more than 27,000 unique single mutations, with over 5000 of them on the S protein, which are intractable for experimental means. {In May 2020},  an intensively validated topology-based neural network tree (TopNetTree) model  \cite{wang2020topology} { was employed to predict certain RBD mutations, including E484K, L452R, and K417N,  would strengthen SARS-CoV-2 infectivity \cite{chen2020mutations}. These predictions have been confirmed \cite{tang2020emergence,mwenda2021detection, faria2021genomic}. Additionally, all 451 new RBD mutations occurred since May 2020 were predicted as the most likely mutations in our work published online last May \cite{chen2020mutations}. We also predicted a list of 625 unlikely RBD mutations \cite{chen2020mutations} and currently, none of them has ever been observed. 
Recently, our TopNetTree model} has been trained on SARS-CoV-2 datasets to accurately predict the S protein and ACE2 or antibody binding free energy changes induced by mutations \cite{chen2020prediction}. A total of 31 disruptive mutations on S protein RBD has been  reported as the potential mutations that { would} most likely disrupt the binding of S protein and essentially all the known SARS-CoV-2 antibodies {had they ever occurred}\cite{chen2020prediction}. Therefore, tracking the growth rate of existing mutations on S protein RBD enables us to monitor the mutations that may impact the efficacy of the existing vaccines and antibody drugs. The study of fast-growing mutations also enables us to understand the SARS-CoV-2 evolutionary tendency and eventually predict future mutations. 

The objective of this work is to track the fast-growing RBD mutations in pandemic-devastated countries and to analyze its evolutionary tendency around the world based on one of the most comprehensive data sets involving 252,874 SARS-CoV-2 genome sequences shown in the Mutation Tracker (\url{https://users.math.msu.edu/users/weig/SARS-CoV-2_Mutation_Tracker.html}). We found  5,420 unique single mutations on the S protein and among them, 825 occurred on the RBD. 
In terms of protein sequence,  95 of 506 non-degenerate mutations on the RBD were observed more than 10 times in the database and are regarded as significant mutations. We show that in addition to 
N501Y, E484K, and K417N, the mutations in the UK, South Africa, and Brazil(ian) variants,  N439K, L452R, S477N,   S477R, and  N501T are also fast-growing mutations in 30 pandemic-devastated countries in the past few months. 
Using the TopNetTree model \cite{wang2020topology,chen2020prediction}, we discover that 88 out of 95 significant mutations on the RBD are associated with the BFE strengthening of the binding of the RBD and ACE2 complex, resulting in more infectious SARS-CoV-2 variants.  {Considering mutation occurrence probability and ability to disrupt antibodies, we identify  vaccine escape and vaccine weakening RBD mutations. }
%Since most RBD mutations were found to have disruptive effects on known antibodies  \cite{chen2020prediction}, 
The present finding suggests that S protein RBD mutations, in general, make the virus more infectious and are disruptive to the existing vaccines and antibody drugs. 

\section{Results}
% RBD: [329,530], and RBM [424,494]

\subsection{Gene-specific analysis on the S protein and the RBD}

Driven by natural selection, random genetic drift, gene editing, host immune responses, etc \cite{sanjuan2016mechanisms,grubaugh2020making}, viruses constantly evolve through mutations, which create genetic diversity and generates new variants.  To have a good understanding of how the mutation will affect the infectivity, transmission, and virulence of SARS-CoV-2, it will be of great importance to study the mutations on SARS-CoV-2, particularly the S protein and the RBD, over a long time period. Therefore, in this work, we mainly focus on the mutations in S protein and S protein RBD. Here, a total of 27,390 unique single mutations has been decoded from 252,874 complete SARS-CoV-2 genome sequences.  

\autoref{tab:Spike gene} shows the distribution of 12 single-nucleotide polymorphism (SNP) types among 5,420 unique mutations and 650,852 non-unique mutations on the S gene of SARS-CoV-2 worldwide. Symbols  N$_{\text{U}}$, N$_{\text{NU}}$, R$_{\text{U}}$, and R$_{\text{NU}}$ represent the number of unique mutations, the number of non-unique mutations, the ratio of 12 SNP types among unique mutation, and the ratio of 12 SNP types among non-unique mutations, respectively. It can be seen that A$>$G and C$>$T have a higher ratio in both unique and non-unique cases, which may be related to the host immune response via APOBEC and ADAR gene editing as reported in \cite{wang2020host}. Moreover, T$>$C has the highest mutation ratios among unique mutations. However, the ratio of T$>$C mutations among the non-unique mutations is not very high, indicating that T$>$C mutations do not commonly occur in the population.

\begin{table}[ht!]
    \centering
    \setlength\tabcolsep{1pt}
	\captionsetup{margin=0.1cm}
	\caption{ The distribution of 12 SNP types among 5,420 unique mutations and 650,852 non-unique mutations on the S gene of SARS-CoV-2 worldwide. N$_{\text{U}}$ is the number of unique mutations and N$_{\text{NU}}$ is the number of non-unique mutations. R$_{\text{U}}$ and R$_{\text{NU}}$ represent the ratios of 12 SNP types among unique and non-unique mutations. In this table, we bold the ratios that are greater than 10\%. }
    \label{tab:Spike gene}
    \begin{tabular}{clcccc|clccccccccc}
    \toprule
    SNP Type & Mutation Type & N$_{\text{U}}$ & N$_{\text{NU}}$ & R$_{\text{U}}$  & R$_{\text{NU}}$ & SNP Type & Mutation Type  & N$_{\text{U}}$ & N$_{\text{NU}}$ & R$_{\text{U}}$  & R$_{\text{NU}}$\\
    \midrule
    A$>$T & Transversion & 489 & 18410 & 9.02\% & 2.83\%  & C$>$T & Transition & 557 & 201705 & {\bf 10.28\%} & {\bf 30.99\%} \\
    A$>$C & Transversion & 387 & 3582 & 7.14\% & 0.55\% & C$>$A & Transversion  & 352 & 36149 & 6.49\% & 5.55\%\\
    A$>$G & Transition   & 771 & 250750 & {\bf 14.23\%} & {\bf 38.53\%} & C$>$G & Transversion  & 169 & 1376 & 3.12\% & 0.21\%\\
    T$>$A & Transversion & 393 & 2071 & 7.25\% & 0.32\% & G$>$T & Transversion  & 446 & 44765 & 8.23\% & 6.88\%\\
    T$>$C & Transition   & 837 & 26874 & {\bf 15.44\%} & 4.13\%  & G$>$C & Transversion  & 242 & 19198 & 4.46\% & 2.95\%\\
    T$>$G & Transversion & 319 & 14358 & 5.89\% & 2.21\%  & G$>$A & Transition    & 458 & 31614 & 8.45\% & 4.86\%\\
    \bottomrule
    \end{tabular}
\end{table}

 \autoref{tab:RBD gene} shows the distribution of 12 SNP types among 825 unique mutations and 50,291 non-unique mutations on the spike RBD gene sequence of SARS-CoV-2 worldwide. To be noticed, compared to \autoref{tab:Spike gene}, the distribution of 12 SNP types acts differently on S protein RBD. The top 3 highest mutation ratios among non-unique mutations are G$>$A, A$>$T, and C$>$A, which indicating that these 3 types of mutations may have a higher impact on the transmission of SARS-CoV-2.

\begin{table}[ht!]
    \centering
    \setlength\tabcolsep{1pt}
	\captionsetup{margin=0.1cm}
	\caption{ The distribution of 12 SNP types among 825 unique mutations and 50,291 non-unique mutations on the spike RBD gene of SARS-CoV-2 worldwide. N$_{\text{U}}$ is the number of unique mutations and N$_{\text{NU}}$ is the number of non-unique mutations. R$_{\text{U}}$ and R$_{\text{NU}}$ represent the ratios of 12 SNP types among unique and non-unique mutations. In this table, we bold the ratios that are greater than 10\%.}
    \label{tab:RBD gene}
    \begin{tabular}{clcccc|clccccccccc}
    \toprule
    SNP Type & Mutation Type & N$_{\text{U}}$ & N$_{\text{NU}}$ & R$_{\text{U}}$  & R$_{\text{NU}}$ & SNP Type & Mutation Type  & N$_{\text{U}}$ & N$_{\text{NU}}$ & R$_{\text{U}}$  & R$_{\text{NU}}$\\
    \midrule
    A$>$T & Transversion & 67 & 12299 & 8.12\% & {\bf 24.46\%}  & C$>$T & Transition & 80 & 4375 & 9.70\% & 8.70\% \\
    A$>$C & Transversion & 56 & 797 & 6.79\% & 1.58\% & C$>$A & Transversion  & 51 & 7287 & 6.18\% & {\bf 14.49\%}\\
    A$>$G & Transition   & 108 & 791 & {\bf 13.09\%} & 1.57\% & C$>$G & Transversion  & 29 & 426 & 3.52\% & 0.85\%\\
    T$>$A & Transversion & 70 & 252 & 8.48\% & 0.50\% & G$>$T & Transversion  & 63 & 2516 & 7.64\% & 5.00\%\\
    T$>$C & Transition   & 140 & 2058 & {\bf 16.97\%} & 4.09\%  & G$>$C & Transversion  & 37 & 232 & 4.48\% & 0.46\%\\
    T$>$G & Transversion & 57 & 1464 & 6.91\% & 2.91\%  & G$>$A & Transition  & 67 & 17794 & 8.12\% & {\bf 35.38\%}\\
    \bottomrule
    \end{tabular}
\end{table}

\begin{figure}[ht!]
    \centering
    \includegraphics[width = 1.0\textwidth]{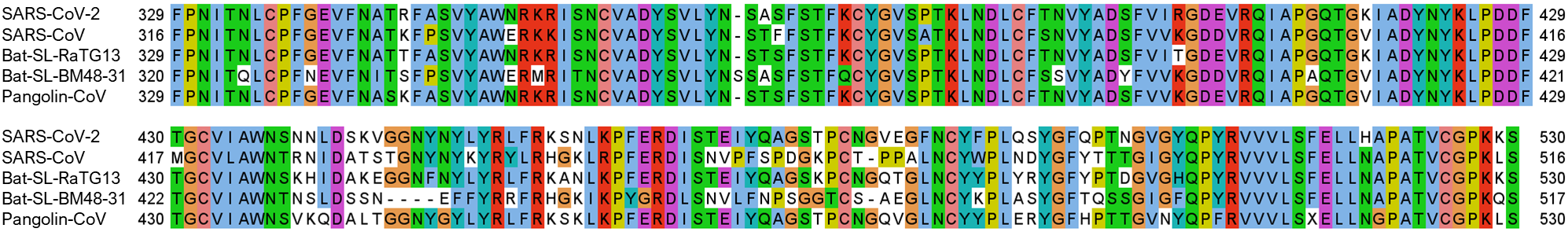}
    \caption{2D sequence alignment for the S protein RBD of SARS-CoV-2, Bat-SL-RaTG13, Pangolin-CoV, SARS-CoV, and Bat-SL-BM48-31.}
    \label{fig:S protein 2D alignment}
\end{figure}

 \autoref{fig:S protein 2D alignment} is the 2D amino acid sequence alignment for the S protein RBD of SARS-CoV-2, Bat-SL-RaTG13, Pangolin-CoV, SARS-CoV, and Bat-SL-BM48-31. It can be seen that residues R346, N354, K417, N438, N440, S443, K444, V445, K458, N460, T478, S494, Q495, and Q498 located on the S protein RBD is not conservative, while the other residues are relatively conservative among different species.

% It can be seen that the S2 domain is relatively conservative than the S1 domain among different species. Notably, the receptor-binding domain (RBD) is located on the S1 domain of S protein, indicating that mutations regulated by natural selection, gene editing, random genetic drift, host immune responses are more likely to occur on the RBD rather than S2 domain of the S protein of SARS-CoV-2. 

\subsection{Impacts of SARS-CoV-2 spike RBD mutations on SARS-CoV-2 infectivity }

\begin{figure}[ht!]
    \centering
    \includegraphics[width = 1.0\textwidth]{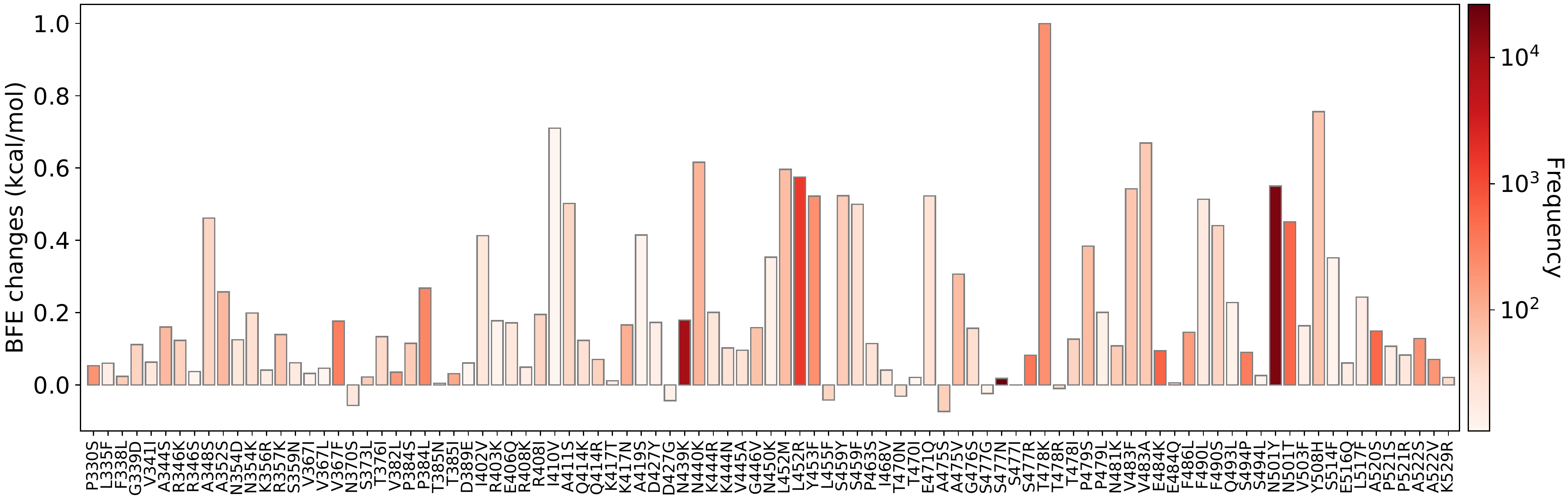}
    \caption{
Illustration of SARS-CoV-2 mutation-induced BFE changes for the complexes of S protein and ACE2. Here, the significant mutations all have frequencies being greater than 10. }
    \label{fig:6M0J}
\end{figure}

The RBD is located on the S1 domain of the S protein, which plays a vital role in binding with the human ACE2 to get entry into host cells. The mutations that are detected on the RBD may affect the binding process and lead to the BFE changes. In this section, we apply the TopNetTree model \cite{chen2020prediction} to predict the mutation-induced BFE changes of RBD and ACE2. \autoref{fig:6M0J} illustrates the predicted BFE changes for S protein and human ACE2 induced by single-site mutations on the RBD. Here, only significant mutations with frequencies being greater than 10 will be considered. The bar plot of mutations with frequencies smaller than 10 can be found in the Supporting Information.  In this figure, a total of 95 significant mutations are displayed. Among them, 7 mutations induced the negative BFE changes, while the other 88 mutations are binding-strengthening mutations. Mutation T478K has the largest BFE changes which are nearly 1 kcal/mol. To be noted, the residue T478 is not conservative among different species as illustrated in \autoref{fig:S protein 2D alignment}. The S477N, N501Y, and N439K mutations are the top 3 significant mutations. Among them, the N501Y mutation has a relatively high BFE change of 0.55 kcal/mol. Moreover, the frequency and predicted BFE changes are both at a high level for mutations L452R, N501T, Y508H. 

\begin{figure}[ht!]
    \centering
    \includegraphics[width = 0.6\textwidth]{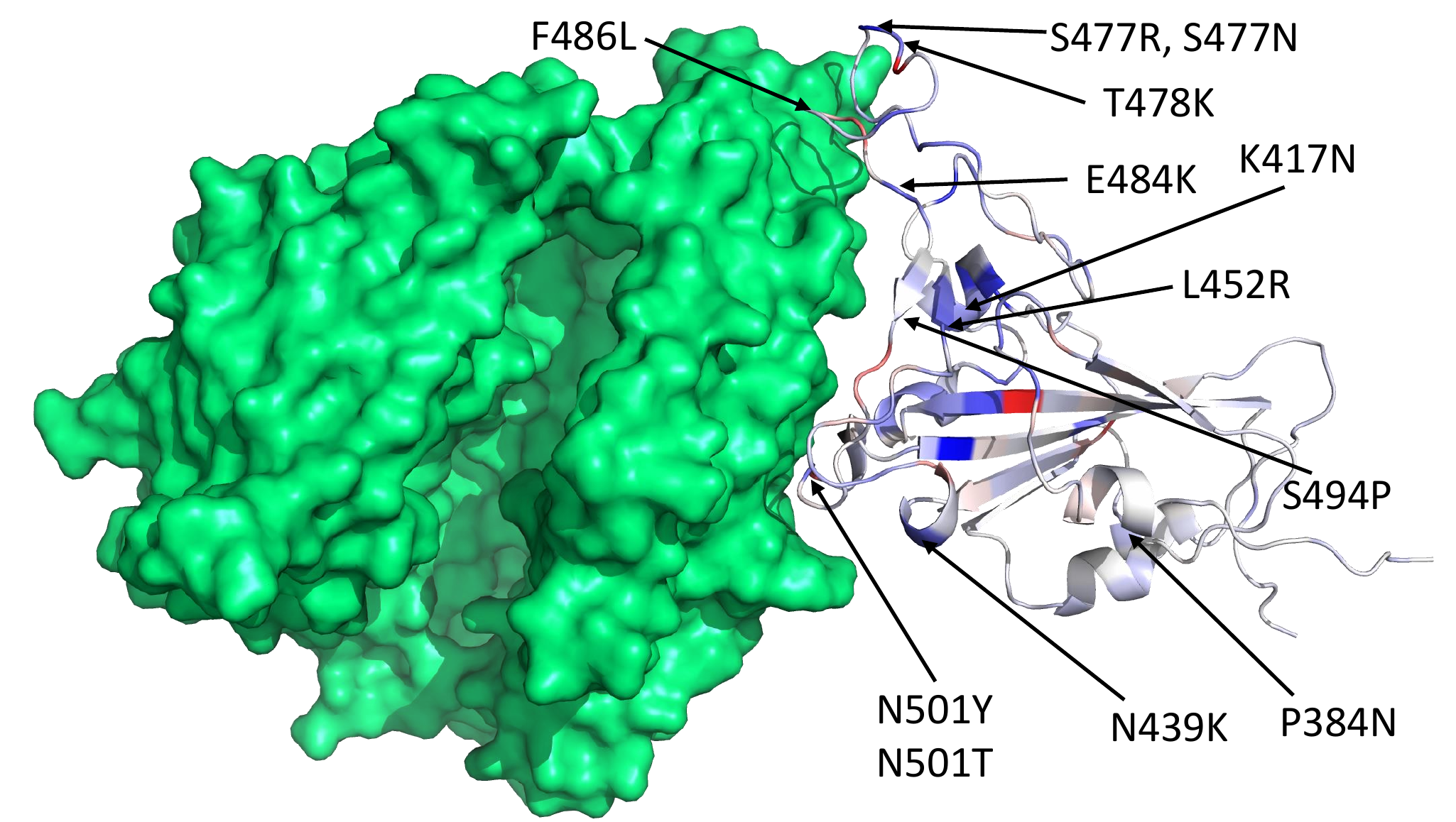}
    \caption{The 3D structure of SARS-CoV-2 S protein RBD bound with ACE2 (PDB ID: 6M0J). We choose blue and red colors to mark the binding-strengthening and binding-weakening mutations, respectively. Vaccine escape mutations described in \autoref{tab:escapemutations} are labeled.}
    \label{fig:6M0J_RBD}
\end{figure}

\autoref{fig:6M0J_RBD} shows the 3D structure of SARS-CoV-2 S protein RBD bound with ACE2. Here, we mark 6 mutations with either high frequency or high BFE changes. The blue and red colors represent the mutations that have positive and negative BFE changes, respectively. The darker the color is, the larger the absolute value of BFE changes is. 

% The blue color represents the positive BFE changes, while  the red color represents  the negative BFE changes. Notably, when the mutant type is proline (P), the BFE changes will more likely have a negative value. This finding consistent with the fact that when proline cannot donate a hydrogen bond to stabilize an $\alpha$-helix, which may destabilize the S protein RBD. 

\subsection{Impacts of SARS-CoV-2 spike RBD mutations on  COVID-19 vaccines}

As reported   early \cite{chen2020prediction}, nearly 71\% mutations on the S protein RDB will weaken the binding of S protein and antibodies, while 64.9\% mutations on the RBD will strengthen the binding of S protein and ACE2, suggesting that these mutations may potentially enhance the infectivity of SARS-CoV-2.   A total of 31 mutations on RBD are reported to significantly weaken the binding of the S protein and most of 51 SARS-CoV-2 antibodies, indicating that these mutations may make the existing vaccine less effective. Such mutations are called the { antibody disrupting} mutations, which are listed in \autoref{tab:escape}. Notably, {most antibody disrupting mutations have negative BFE changes, suggesting that they will make the SARS-CoV-2 less infectious and thus, will not frequently occur due to the natural selection. As a result, many of them may not be able to evade the existing vaccines in a population}. 
\begin{table}[ht!]
    \centering
    \setlength\tabcolsep{5pt}
	\captionsetup{margin=0.1cm}
	\caption{The most antibody disruptive (AD) RBD mutations and their corresponding BFE changes (unit: kcal/mol) of the binding of S protein and ACE2. }
    \label{tab:escape}
    \begin{tabular}{cc|cc|cccccccccccc}
    \toprule
    AD Mutation & BFE changes & AD Mutation & BFE changes  & AD Mutation & BFE changes\\
    \midrule
    E406G & 0.4908 & I418N & -2.7394 & N422K & -2.3917 \\
    D442H & -0.9957 & Y505S & -1.9337 & Y421D & -1.2569\\
    R355W & -0.9873 & F400I & -2.5435 &  F400C & -2.3813\\
    I402F & -2.8545 &  C432G & -2.7148 & I434K & -0.9279\\
    A435P & -2.4472 &  Q493P & 0.0444 &  V510E & -2.7094\\
    V512G & -2.793 &  L513P & -2.8153 & V350F & -2.5921\\
    W353R & -0.7866 & I410N & -2.8721 &  G416V & -2.7848 \\
    G431V & -2.3621 &  Y449D & -1.0226 & Y449S & -0.8112\\
    L461H & -0.7456 &  S469P & -1.2212 &  C480R & -2.4793\\
    P491R & -3.1152  & P491L & -2.184 & Y495C & -1.4773 \\
    Q506P & -3.5875 & & & &\\
    \bottomrule
    \end{tabular}
\end{table}

{We hypothesize that RBD mutations that can simultaneously strengthen the infectivity and disrupt the binding between the S protein and existing antibodies will pose imminent threats to the current crop of vaccines. In other word, vaccine escape (VE) mutations are both fast-growing and antibody disrupting.  We also define vaccine weakening (VW) as those fast-growing mutations that will moderately weaken the binding of the S protein and many existing antibodies. Based on the fast-growing RBD mutations detected since the beginning of 2021, we predict a list of vaccine escape, vaccine weakening RBD mutations in \autoref{tab:escapemutations}. Fast growing RBD mutation that do not significantly weaken most antibody bindings are presented in  \autoref{tab:escapemutations}. It is of great importance to track not only the ACE2-binding-strengthening RBD mutations but also the antibody-binding-weakening RBD mutations. }

{
\begin{table}[ht!]
    \centering
    \setlength\tabcolsep{5pt}
	\captionsetup{margin=0.1cm}
	\caption{List of vaccine escape (VE), vaccine weakening (VW), and  fast-growing (FG) mutations. 
	Their corresponding BFE changes (unit: kcal/mol) of the binding of S protein and ACE2 are provided as well. }
    \label{tab:escapemutations}
    \begin{tabular}{cc|cc|cccccccccccc}
    \toprule
    VE Mutation & BFE changes & VW Mutation & BFE changes  & FG Mutation & BFE changes\\
    \midrule
N501Y & 0.5499	 &    S477N & 0.018	 &   A520S & 0.1495\\	
L452R & 0.5752	 &    N439K & 0.1792 &   T385I & 0.0314	\\
E484K & 0.0946   &    S477R & 0.082	 &   A522S & 0.1283	\\
N501T & 0.4514	 &    V367F & 0.1764 &   N440K & 0.6161	 \\
S494P & 0.0902	 &    Q414R & 0.0708 &   A352S & 0.2576	\\
T478K & 0.9994	 &    T470N & -0.031 &   V382L & 0.0355	\\
K417N & 0.1661	 &          &        &   P479S & 0.3844	 \\   
F486L & 0.1456	 &          &        &   A522V & 0.0705	\\
P384L & 0.2681	 &          &        &   S459Y & 0.5234	\\
P384S & 0.1151   &          &        &   G339D & 0.1117\\
K417T & 0.0116   &          &        &         &        \\  
    \bottomrule
    \end{tabular}
\end{table}
}

\subsection{Fast-growing mutations in COVID-19-devastated countries}

In this section, we extract the 30 countries with the highest number of SNP profiles and analyze their mutations on S protein RBD, as illustrated in \autoref{tab:Countries}. We can see that the BFE changes of S protein and ACE2 induced by mutations on the RBD are mostly positive, suggesting that the binding of ACE2 and S protein will be potentially strengthened in these 30 countries. This indicates that SARS-CoV-2 becomes more infectious, driven by most mutations on the receptor-binding domain. 

\begin{table}[ht!]
    \centering
    \setlength\tabcolsep{15pt}
	\captionsetup{margin=0.1cm}
	\caption{The statistical analysis of mutations on S protein RBD of 30 countries with large sequencing data. N$_{\text{seq}}$ is the number of sequences in each country. N$_{\text{U-RBD}}$ is the number of unique mutations on RBD and N$_{\text{NU-RBD}}$ is the number of non-unique mutations on RBD. N$_{\text{positive}}$ and N$_{\text{negative}}$ represent the number of unique single mutations that will respectively result in positive and negative BFE changes of S protein and ACE2 induced by mutations on S protein RBD.}
    \label{tab:Countries}
    \begin{tabular}{lcccccclccccccccc}
    \toprule
    Country  & N$_{\text{seq}}$ & N$_{\text{U}}$ & N$_{\text{NU}}$ & N$_{\text{positive}}$ & N$_{\text{negative}}$   \\
    \midrule
    United Kingdom & 90972 & 201 & 15158 & 162 & 39 \\
    USA & 55063 & 187 & 2834 & 149 & 38 \\
    Denmark & 25097 & 84 & 7181 & 74 & 10 \\
    Australia & 9583 & 37 & 7462 & 30 & 7 \\
    Canada & 9504 & 42 & 165 & 36 & 6 \\
    Netherlands & 5232 & 41 & 1211 & 39 & 2 \\
    Switzerland & 4922 & 40 & 1525 & 39 & 1 \\
    France & 3397 & 36 & 1205 & 32 & 4 \\
    Iceland & 3119 & 13 & 158 & 13 & 0 \\
    India & 3046 & 37 & 71 & 32 & 5 \\
    Belgium & 2903 & 37 & 537 & 37 & 0 \\
    Luxembourg & 2594 & 23 & 1058 & 22 & 1 \\
    Spain & 2375 & 36 & 166 & 30 & 6 \\
    Germany & 2345 & 26 & 230 & 25 & 1 \\
    Italy & 2219 & 29 & 296 & 26 & 3 \\
    United Arab Emirates & 1581 & 21 & 80 & 21 & 0 \\
    Sweden & 1302 & 18 & 326 & 18 & 0 \\
    Singapore & 1268 & 16 & 66 & 15 & 1 \\
    Brazil & 1244 & 14 & 207 & 12 & 2 \\
    Russia & 1060 & 22 & 68 & 20 & 2 \\ 
    Norway & 1021 & 13 & 196 & 13 & 0 \\
    Portugal & 947 & 14 & 90 & 13 & 1 \\
    Chile & 888 & 2 & 2 & 2 & 0 \\
    Ireland & 877 & 11 & 191 & 11 & 0 \\
    South Africa & 851 & 20 & 84 & 17 & 3 \\
    Japan & 713 & 2 & 2 & 2 & 0 \\
    Austria & 705 & 10 & 84 & 9 & 1 \\
    Israel & 693 & 19 & 117 & 19 & 0 \\
    Mexico & 593 & 10 & 83 & 10 & 0 \\
		China & 565 & 6 & 12 & 5 & 1\\
    \bottomrule
    \end{tabular}
\end{table}

\begin{figure}[ht!]
    \centering
    \includegraphics[width = 1\textwidth]{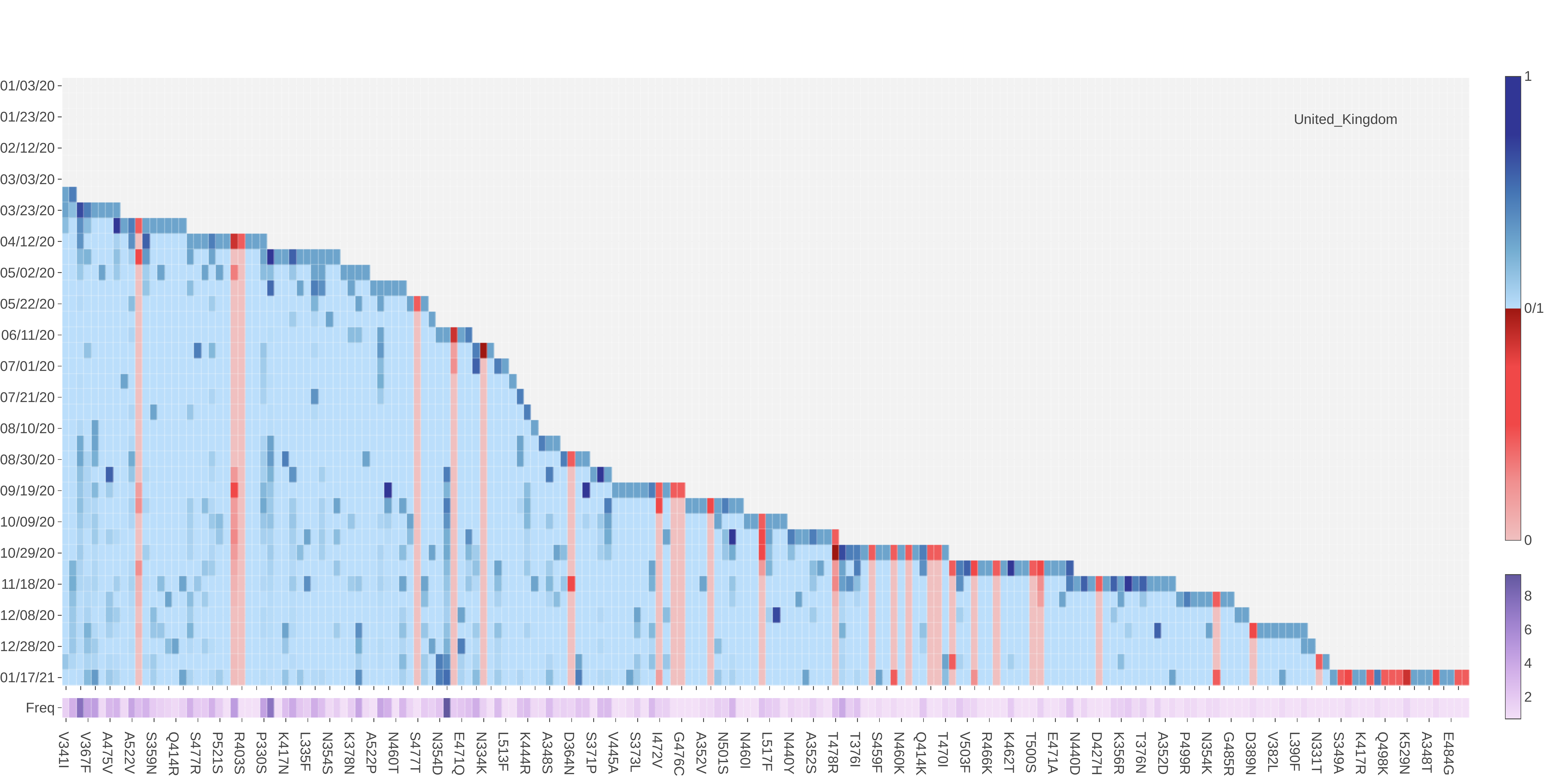}
    \caption{The log growth ratio and log frequency of mutations on S protein RBD in the United Kingdom. The blue and red colors respectively represent the binding-strengthening and binding-weakening mutations on RBD. The darker blue/red means the binding-strengthening/binding-weakening mutations with a higher growth ratio in a specific 10-day period. The darker purple represents the mutation with a higher log frequency. }
    \label{fig:UK}
\end{figure}

Tracking the binding-strengthening mutations will play a vital role in the development of anti-virus drugs, antibody drugs, and vaccines. Therefore, we calculate the growth ratio of mutations on the RBD  on a 10-day average, aiming to monitor the binding-strengthening mutations that have rapid growth over time. \autoref{fig:UK} illustrates the log growth ratio and log frequency of mutations on the S protein RBD in the United Kingdom on a 10-day average. The blue and red colors respectively represent the positive and negative BFE changes induced by a specific mutation, and the purple color represents the log frequency of a specific mutation. The darker the color is, the higher the log growth ratio/log frequency will be. For a better view, please check the HTML file in our Supporting Information. From \autoref{fig:UK}, we can see that the N501Y mutation with a positive BFE change have a relatively high growth ratio since early September 2020, which consist with the news that a new strain B.1.1.7 (also known as 20I/501Y.V1) in the United Kingdom has the potential to increase the pandemic trajectory \cite{galloway2021emergence}. Moreover, mutations V367F, E484K, N355D, and S373L  with positive BFE changes also have a relatively higher mutation ratio since early 2021, indicating that these four mutations may strengthen the binding of ACE2 and the S protein RBD, and potentially increase the infectivity of SARS-CoV-2. Reported in Ref. \cite{chen2020prediction}, mutation E484K may dramatically disruptive effects on antibodies B38, CV30, Sb23, Fabs 298 52, and CV30. To be noted, since early 2021, the number of binding-weakening mutations has an increasing tendency, such as the S394A, S477G, G477D, F456L, K529N, R408T, and G477V.

\begin{figure}[ht!]
    \centering
    \includegraphics[width = 1\textwidth]{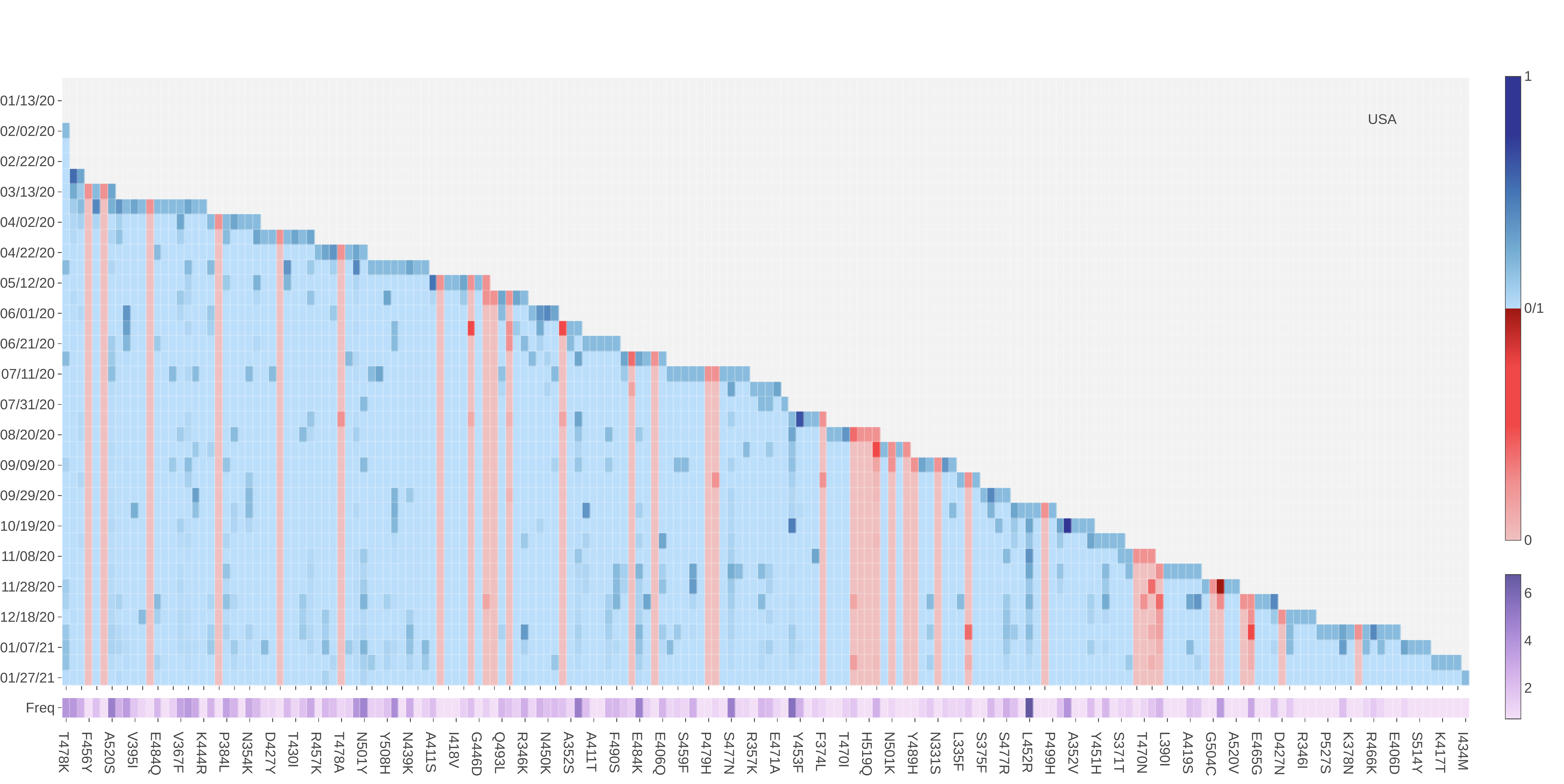}
    \caption{The log growth ratio and log frequency of mutations on S protein RBD in the United States. The blue and red colors respectively represent the binding-strengthening and binding-weakening mutations on RBD. The darker blue/red means the binding-strengthening/binding-weakening mutations with a higher growth ratio in a specific 10-day period.  The darker purple represents the mutation with a higher log frequency. }
    \label{fig:USA}
\end{figure}

\autoref{fig:USA} illustrates the log growth ratio and log frequency of mutations on S protein RBD in the United States on a 10-day average. Similar to the United Kingdom, the N501Y, E484K recently have a high log growth rate. Additionally, the binding-strengthening mutations T385I, N439K, S477R, and L452R also have a high log growth ratio since late 2020. To be noted, L452R had been reported as the key mutation that linked to COVID-19 outbreaks in California on January 17, 2021 \cite{zhang2021emergence}.

\begin{figure}[ht!]
    \centering
    \includegraphics[width = 1\textwidth]{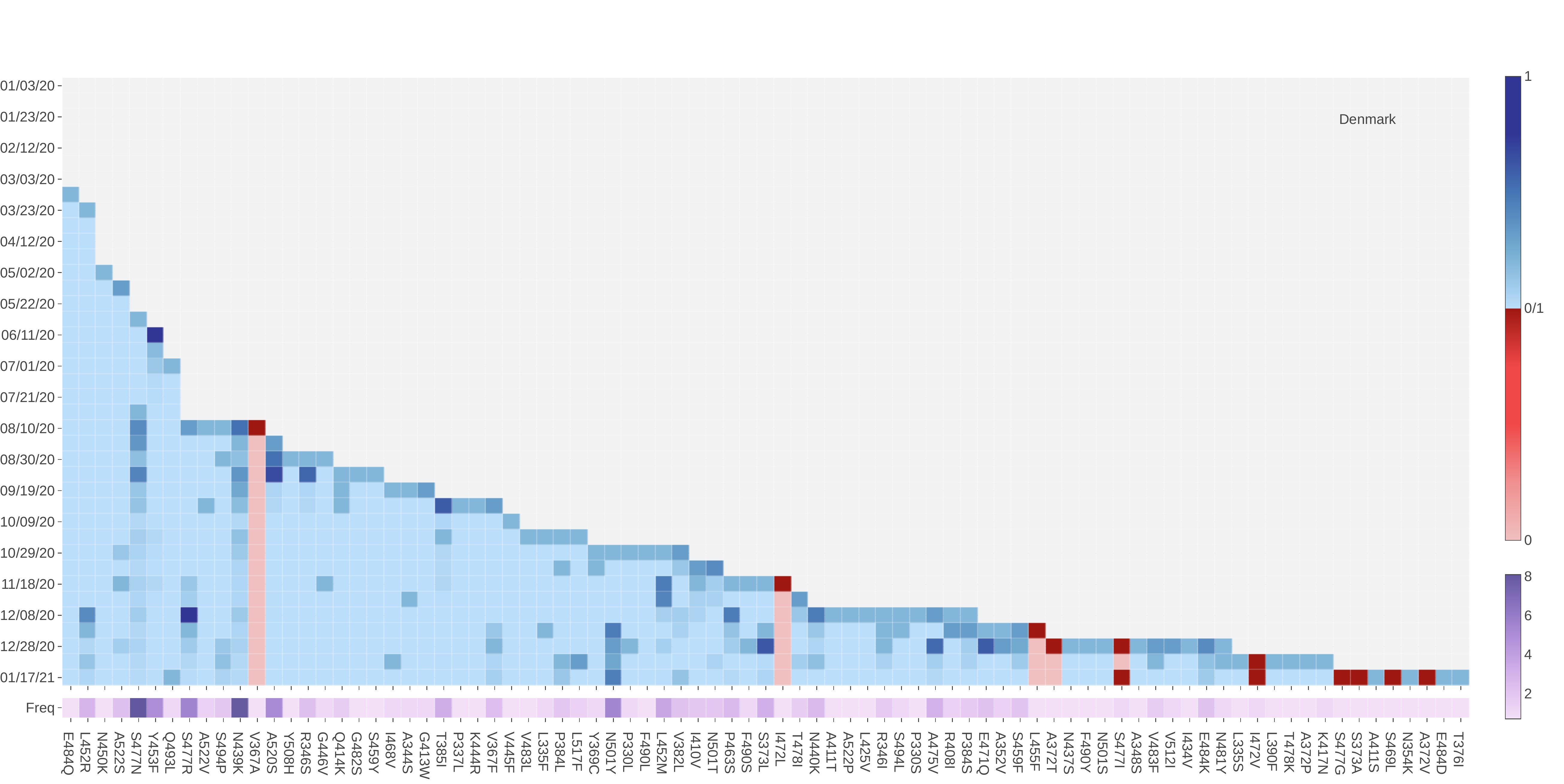}
    \caption{The log growth ratio and log frequency of mutations on S protein RBD in the Denmark. The blue and red colors respectively represent the binding-strengthening and binding-weakening mutations on RBD. The darker blue/red means the binding-strengthening/binding-weakening mutations with a higher growth ratio in a specific 10-day period.  The darker purple represents the mutation with a higher log frequency.}
    \label{fig:Denmark}
\end{figure}

\begin{figure}[ht!]
    \centering
    \includegraphics[width = 1\textwidth]{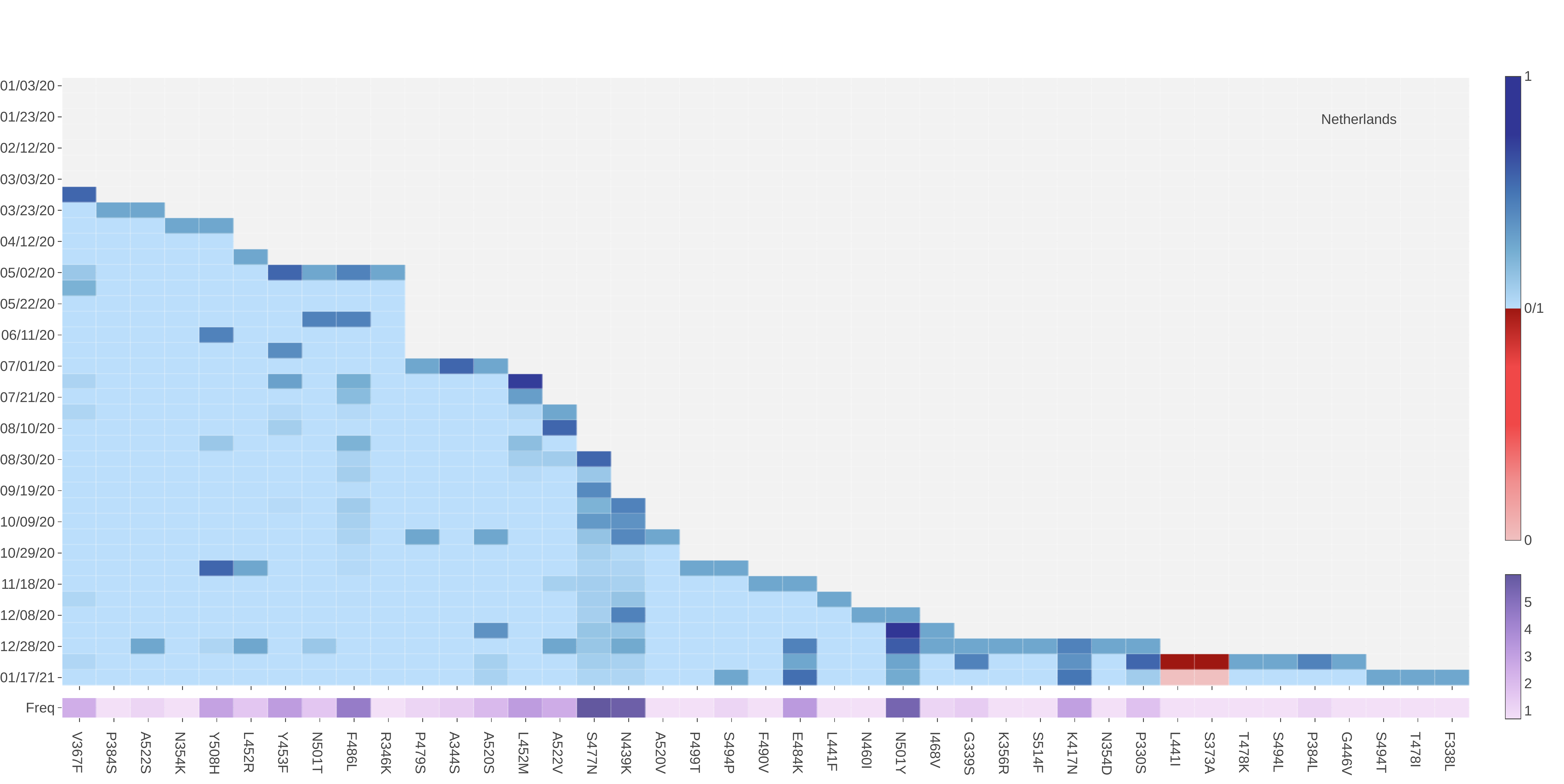}
    \caption{ The log growth ratio and log frequency of mutations on S protein RBD in the Netherlands. The blue and red colors respectively represent the binding-strengthening and binding-weakening mutations on RBD. The darker blue/red means the binding-strengthening/binding-weakening mutations with a higher growth ratio in a specific 10-day period.  The darker purple represents the mutation with a higher log frequency.}
    \label{fig:Netherlands}
\end{figure}

\autoref{fig:Denmark} tracks the fast-growing mutations in Denmark. Binding-strengthening mutation L452R has a fast-growing tendency since December 8, 2020. Binding-strengthening mutation S477N has a high growth ratio from late July to early December. Mutation S477R that induced the positive BFE changes has a very rapid growth between November 28, 2020, to December 08, 2020, while the number of S447R mutations has recently not increased rapidly.  The number of the binding-strengthening mutation N439K keeps a high growth rate since early August. However, the increasing rate of the N439K mutation slows down recently.  As first reported in the United Kingdom, the N501Y mutation also has a fast-growing tendency since early December 2020, making the SARS-CoV-2 more infectious. A similar pattern can also be observed in Netherlands, Switzerland, Norway, and Sweden. Moreover, as shown in \autoref{fig:Netherlands}, three binding-strengthening mutations have a rapid growth since late December 2020: V367F, T478K, and P479S. Scientists and researchers worldwide should keep tracking these three mutations in the following months.

\begin{figure}[ht!]
    \centering
    \includegraphics[width = 1\textwidth]{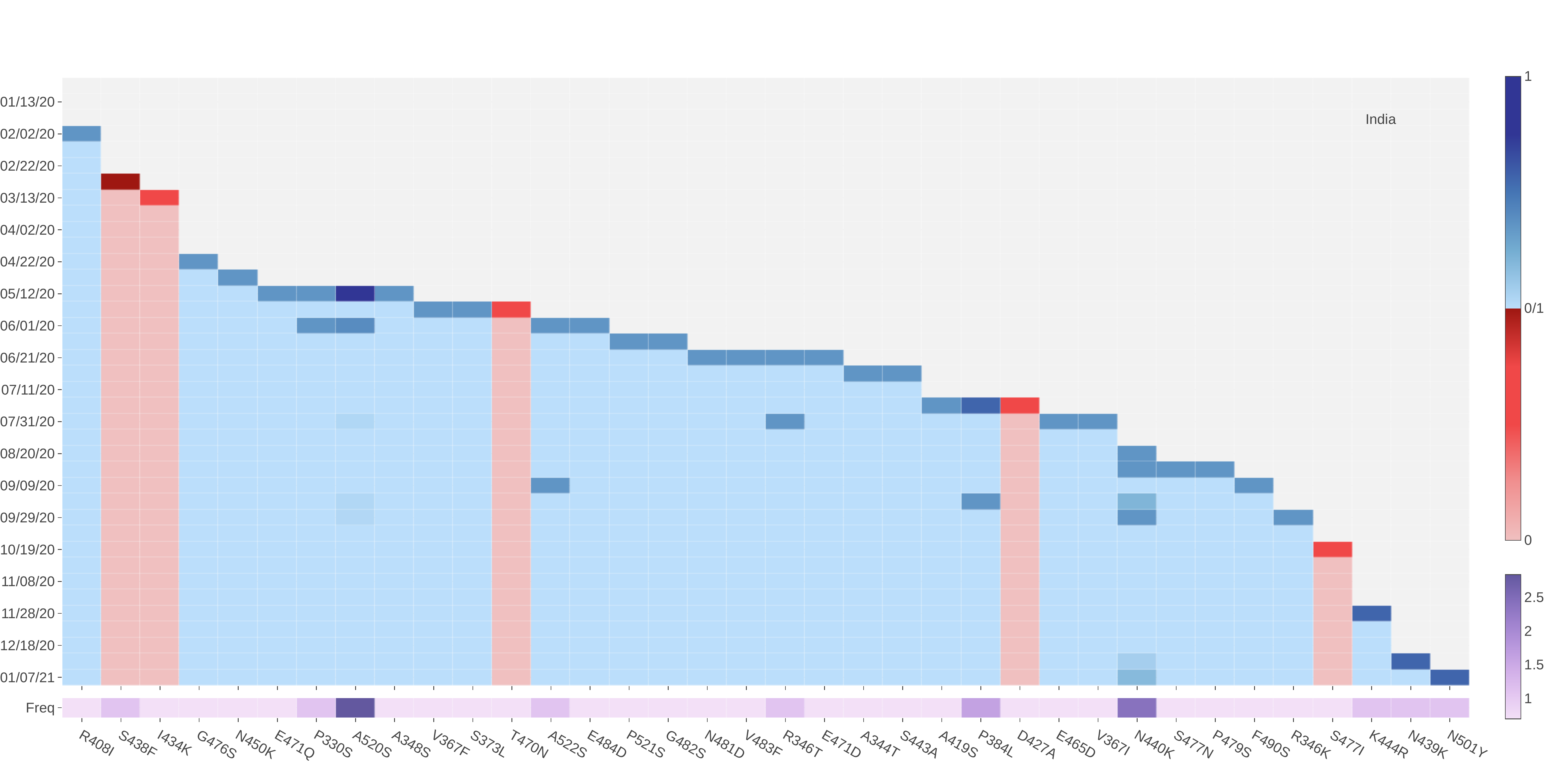}
    \caption{The log growth ratio and log frequency of mutations on S protein RBD in India. The blue and red colors respectively represent the binding-strengthening and binding-weakening mutations on RBD. The darker blue/red means the binding-strengthening/binding-weakening mutations with a higher growth ratio in a specific 10-day period.  The darker purple represents the mutation with a higher log frequency.}
    \label{fig:India}
\end{figure}

\begin{figure}[ht!]
    \centering
    \includegraphics[width = 1\textwidth]{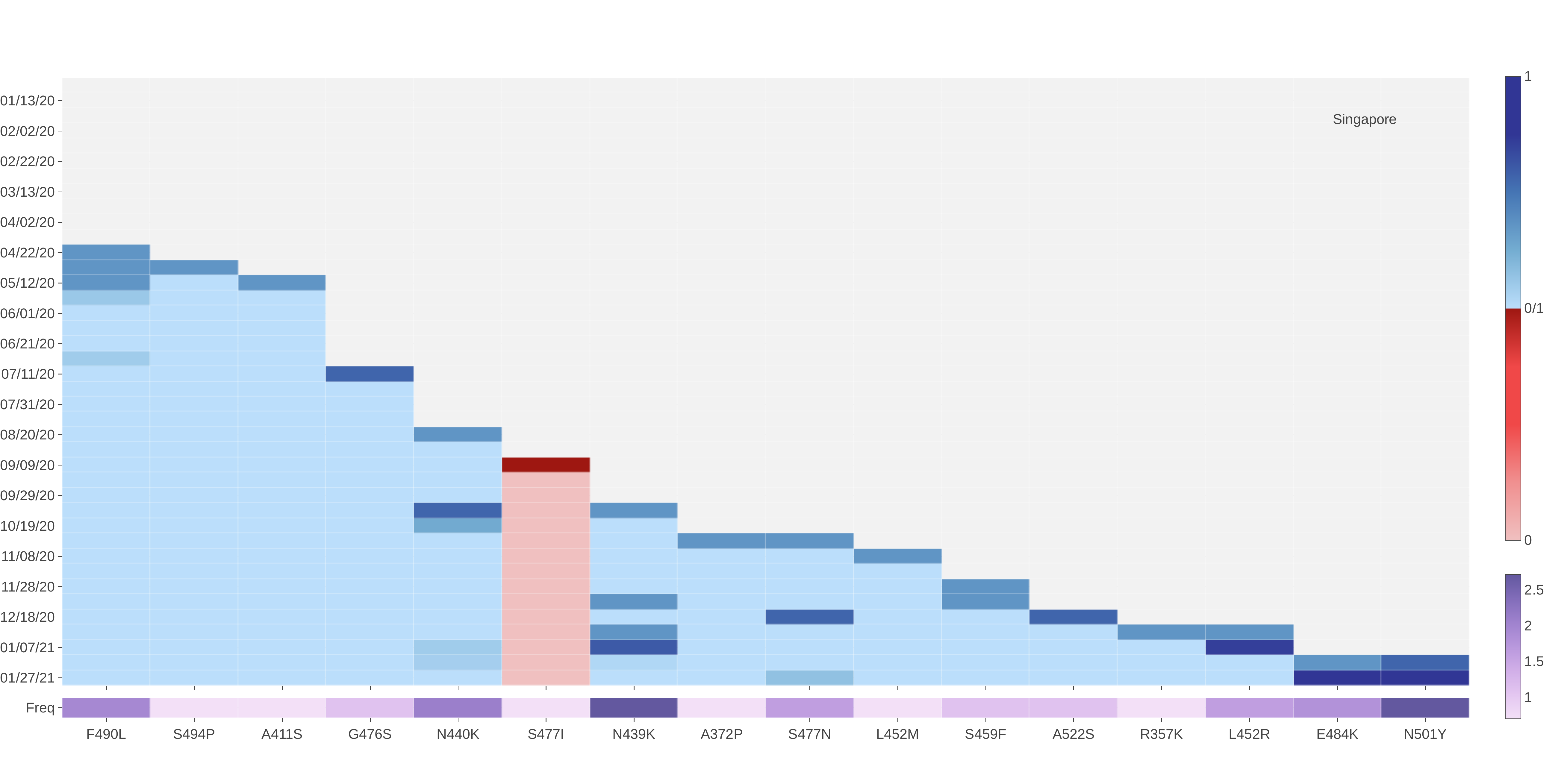}
    \caption{The log growth ratio and log frequency of mutations on S protein RBD in Singapore. The blue and red colors respectively represent the binding-strengthening and binding-weakening mutations on RBD. The darker blue/red means the binding-strengthening/binding-weakening mutations with a higher growth ratio in a specific 10-day period.  The darker purple represents the mutation with a higher log frequency.}
    \label{fig:Singapore}
\end{figure}

Unlike the mutations in the United Kingdom, United States, and Denmark, the only binding-strengthening mutation in India is N440K, which has a relatively high frequency. Although the A530S mutation introduces the positive BFE changes with the highest frequency, the growth rate quite low after early October 2020 (See \autoref{fig:India}).  Singapore also has the binding-strengthening mutations E484K, N501Y, S477N, and L452R, as those found in other countries. Moreover, one binding-strengthening mutation N440K with a high frequency has a relatively high growth rate since 2021 (See \autoref{fig:Singapore}).

\begin{figure}[ht!]
    \centering
    \includegraphics[width = 1\textwidth]{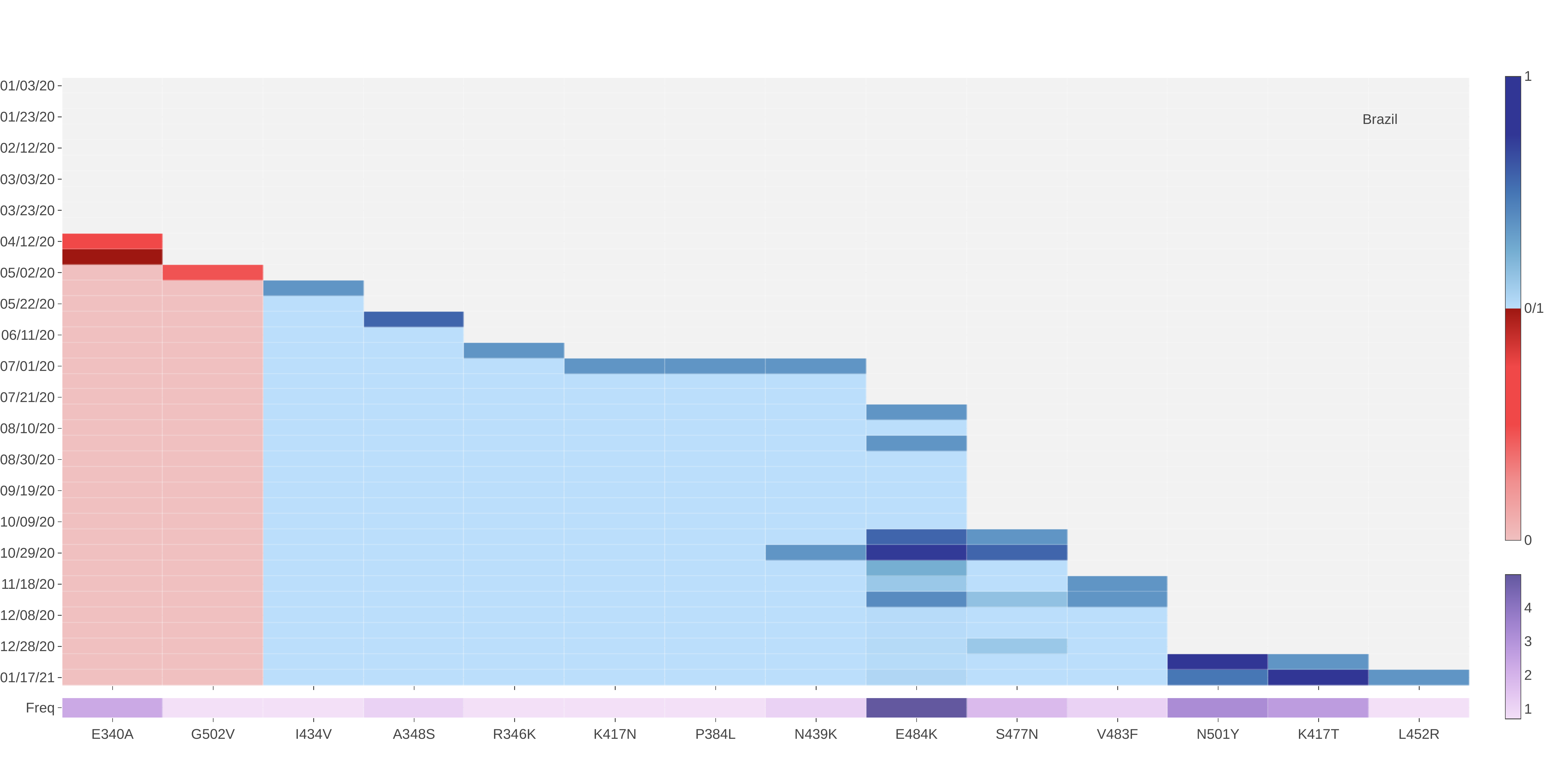}
    \caption{The log growth ratio and log frequency of mutations on S protein RBD in Brazil. The blue and red colors respectively represent the binding-strengthening and binding-weakening mutations on RBD. The darker blue/red means the binding-strengthening/binding-weakening mutations with a higher growth ratio in a specific 10-day period.  The darker purple represents the mutation with a higher log frequency.}
    \label{fig:Brazil}
\end{figure}

\begin{figure}[ht!]
    \centering
    \includegraphics[width = 1\textwidth]{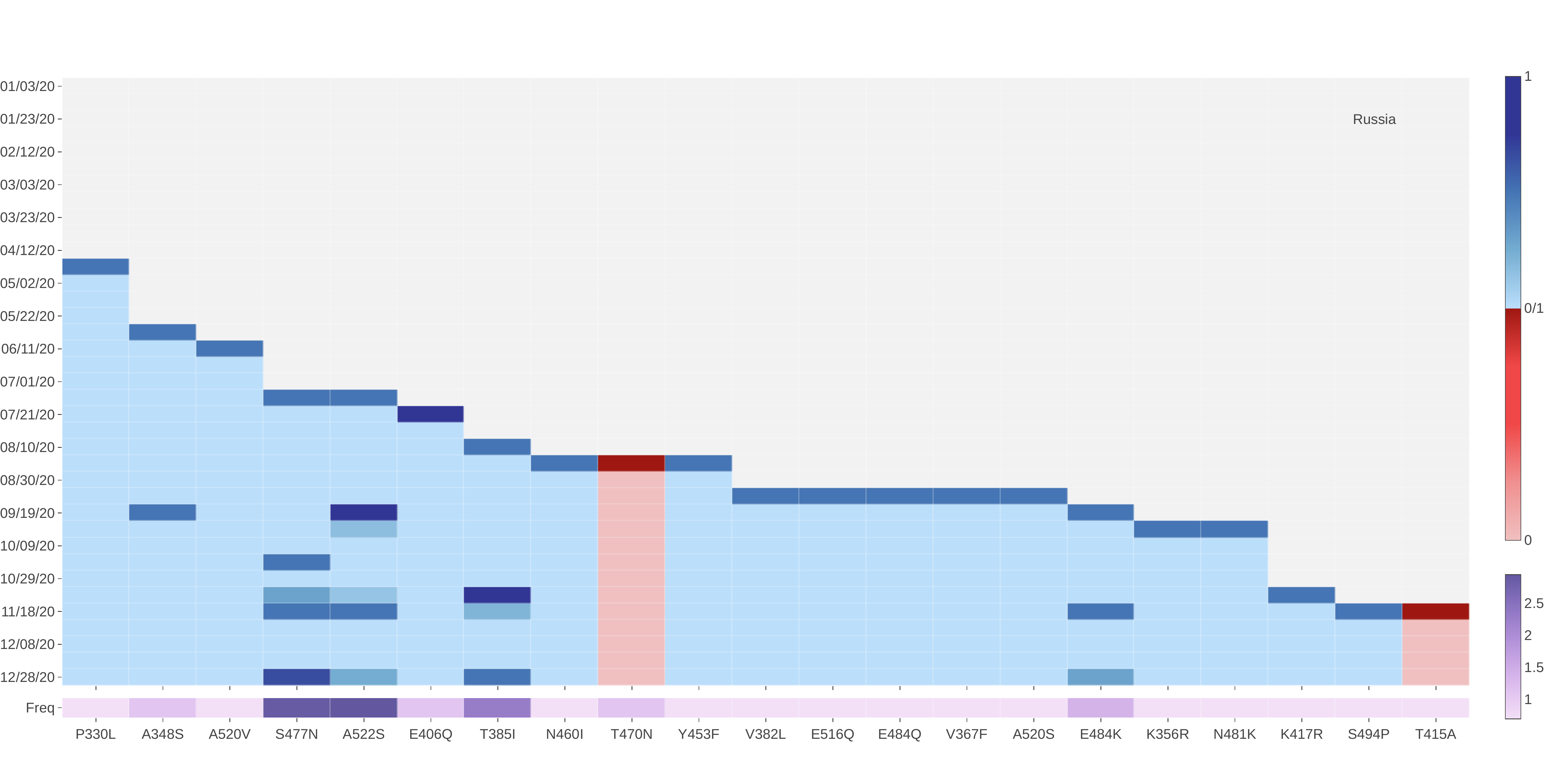}
    \caption{The log growth ratio and log frequency of mutations on S protein RBD in Russia. The blue and red colors respectively represent the binding-strengthening and binding-weakening mutations on RBD. The darker blue/red means the binding-strengthening/binding-weakening mutations with a higher growth ratio in a specific 10-day period.  The darker purple represents the mutation with a higher log frequency.}
    \label{fig:Russia}
\end{figure}

First reported by the National Institute of Infectious Diseases (NIID) in Japan that four travelers from Brazil sampled a branch of the B.1.1.28 lineage called P.1 variant (also known as 20J/501Y.V3) \cite{naveca2021sars}. This variant contains three mutations in the S protein RBD: K417T, E484K, and N501Y. All of them are all the binding-strengthening mutations with a fast growth rate since late December 2020, as illustrated in \autoref{fig:Brazil}. The binding-strengthening mutations in Russia are S477N, A522S, T385I, and E484K. All of them have a high frequency and a fast growth ratio since September 2020 (See \autoref{fig:Russia}).

\begin{figure}[ht!]
    \centering
    \includegraphics[width = 1\textwidth]{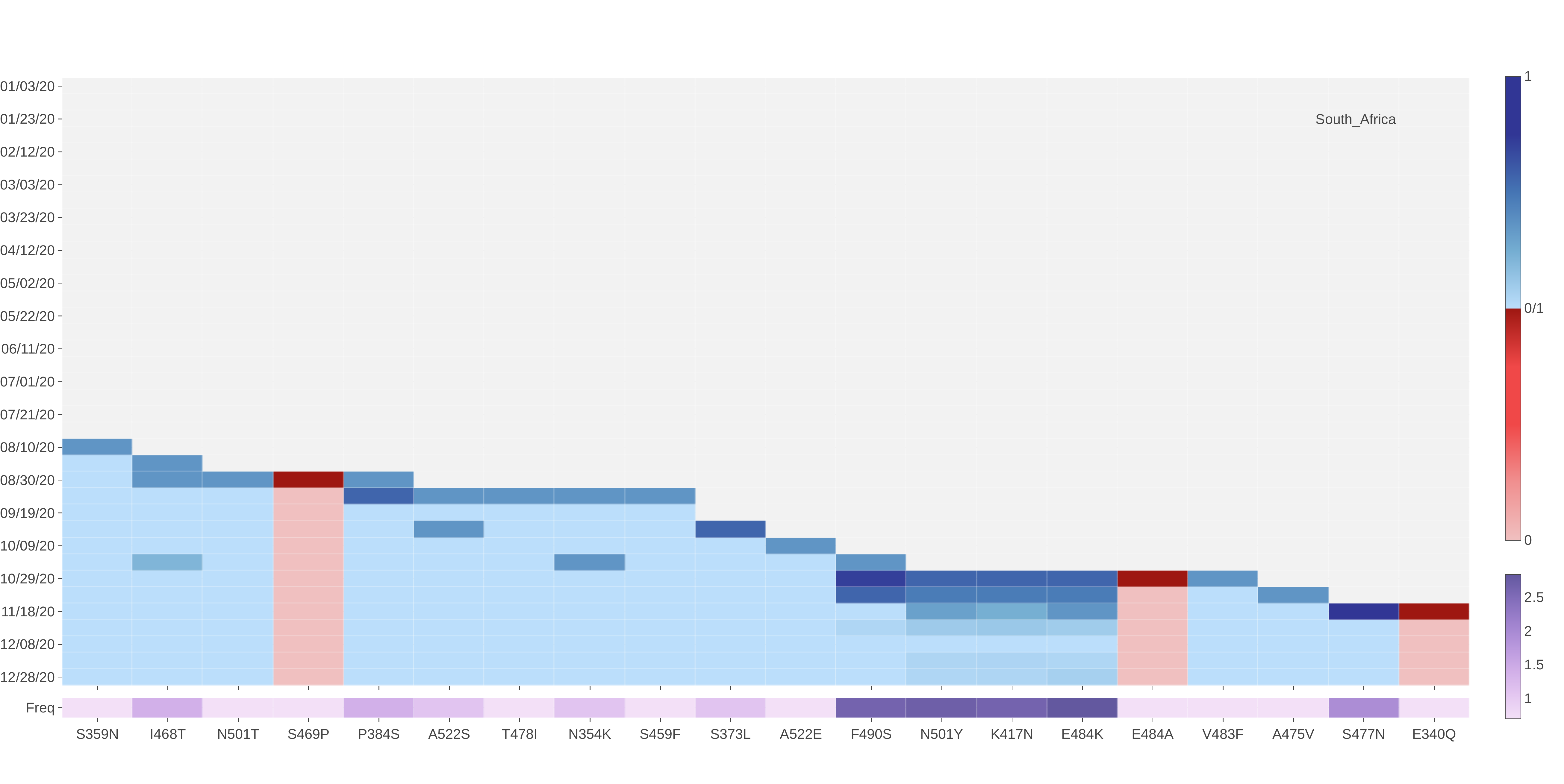}
    \caption{The log growth ratio and log frequency of mutations on S protein RBD in South Africa. The blue and red colors respectively represent the binding-strengthening and binding-weakening mutations on RBD. The darker blue/red means the binding-strengthening/binding-weakening mutations with a higher growth ratio in a specific 10-day period.  The darker purple represents the mutation with a higher log frequency.}
    \label{fig:South Africa}
\end{figure}

 The B.1.351 lineage (also known as 20H/501Y.V2) first identified in Nelson Mandela Bay, South Africa, which can be traced back to the beginning of October 2020, has become a predominant variant in South Africa. From \autoref{fig:South Africa}, we can see that mutations F480S, N501Y, K417N, and E484K have a rapid growing tendency since the beginning of October 2020. Moreover, these four mutations are all the binding-strengthening mutations with a very high frequency, which consistent with the finding of the B.1.351 lineage. This indicates that the predicted BFE changes of S protein and ACE2 from our TopNetTree model are reliable. 

 From analyzing the SNP profiles in Mexico, we notice that 6 binding-strengthening mutations have a rapid growth since late October 2020. They are L452R, S477N, T478K, S494P, E484K, and A552V. Among them, T478K has the highest growth ratio since late October 2020, indicating that T478K may potentially make the SARS-CoV-2 more transmissible and infectious. 

\begin{figure}[ht!]
    \centering
    \includegraphics[width = 1\textwidth]{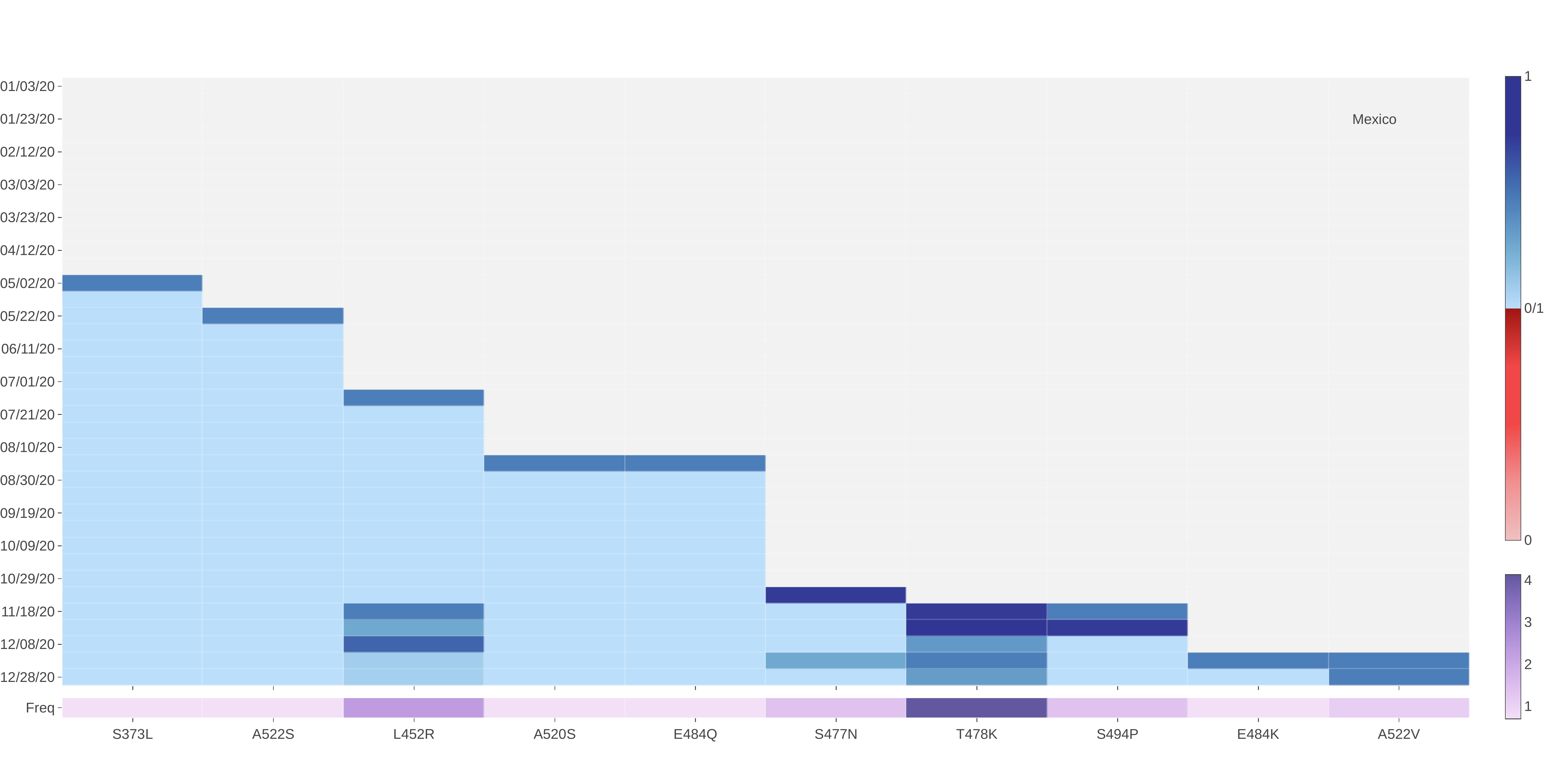}
    \caption{The log growth ratio and log frequency of mutations on S protein RBD in Mexico. The blue and red colors respectively represent the binding-strengthening and binding-weakening mutations on RBD. The darker blue/red means the binding-strengthening/binding-weakening mutations with a higher growth ratio in a specific 10-day period.  The darker purple represents the mutation with a higher log frequency.}
    \label{fig:Mexico}
\end{figure}

\subsection{Discussion}

The BFE changes following 506 non-degenerate mutations on the S protein RBD are presented in Figures S1-S5 of the Supporting information. These plots highlight the magnitude disparity in BFE changes induced by binding-strengthening mutations and binding-weakening mutations. Such a large disparity indicates that SARS-CoV-2 is evolutionarily quite advance with respect to human infection. 
\begin{table}[ht!]
    \centering
    \setlength\tabcolsep{15pt}
	\captionsetup{margin=0.1cm}
	\caption{Most significant mutations on S protein RBD of 30 countries with large sequencing data. }
    \label{tab:Countries_Mutations}
    \begin{tabular}{ll}
    \toprule
    Country  &  Most significant mutations   \\
    \midrule
    United Kingdom &  N439K, S477N, S494P, and N501Y,              \\
    USA &   A520S, N501Y, S494P, E484K, S477N, N501T, and L452R\\
    Denmark & S477N, Y453F, S477R, N439K, and N501Y\\
    Australia &   S477N and N501T  \\
    Canada &   R357K, E484K, and L452R \\
    Netherlands &   F486L, S477N, N439K, and N501Y\\
    Switzerland &   N439K, S477N, N501Y, and T478K\\
    France &   S477N, N501Y, and N501T\\
    Iceland &  S477N and N439K \\
    India &  A520S, P384L, and N440K\\
    Belgium &   S477N, N439K, and N501Y \\
    Luxembourg &   S477N,  N439K, and N501Y\\
    Spain &   S477N and N501Y\\
    Germany &   N439K, S477N, and N501Y\\
    Italy &   N439K, S477N, L452R, and N501Y\\
    United Arab Emirates &   N501Y, N439K, E484K, and K417N\\
    Sweden &    S477N, N439K, and N501Y\\
    Singapore &  F490L, N440K, N439K, E484K, and N501Y \\
    Brazil &   E484K, N501Y, and K417T \\
    Russia &   S477N, A522S, T385I, and E484K\\ 
    Norway &   N439K, S477N, A520S, and N501Y\\
    Portugal &   S477N, L452R, and N501Y\\
    Chile &   P479S and S373L\\
    Ireland &   N439K and N501Y\\
    South Africa &  F490S, N501Y, K417N, E484K, and S477N \\
    Japan &  N394Y and S359N \\
    Austria &   S477N and N439K\\
    Israel &   N481K, N501Y, N439K, and L425R\\
    Mexico &   L452R and T478K \\
		China & P521R   and S477N \\
    \bottomrule
    \end{tabular}
\end{table}
Figures S20-S25 of the Supporting information provide the log growth ratio and log frequency of mutations on S protein RBD in the Australia, Austria, Belgium, Canada, Chile, China, France, Germany, Iceland, Ireland, Israel, Italy, Japan, Luxembourg, Norway, Portugal, Spain, Sweden, Switzerland, and the United Arab Emirates. 
The most significant mutations in Australia are S477N and N501T. 
The most significant mutations in Austria are S477N and N439K. 
The most significant mutations in Belgium are S477N, N439K, and N501Y. 
The most significant mutations in Canada are R357K, E484K, and L452R. 
The most significant mutations in France are S477N, N501Y, and N501T.  
The most significant mutations in Germany are N439K, S477N, and N501Y. 
The most significant mutations in Iceland are S477N and N439K. 
The most significant mutations in Ireland are N439K and N501Y. 
The most significant mutations in Israel are N481K, N501Y, N439K, and L425R. 
The most significant mutations in Italy are N439K, S477N, L452R, and N501Y. 
The most significant mutations in Luxembourg are S477N,  N439K, and N501Y. 
The most significant mutations in Norway are N439K, S477N,A520S, and N501Y. 
The most significant mutations in Portugal are S477N, L452R, and N501Y. 
The most significant mutations in Spain are S477N and N501Y. 
The most significant mutations in Sweden are S477N, N439K, and N501Y. 
The most significant mutations in Switzerland are N439K, S477N, N501Y, and T478K. 
The most significant mutations in the United Arab Emirates are N501Y, N439K, E484K, and K417N. 
This information, together with those given in Figures \ref{fig:UK}-\ref{fig:Mexico}, shows that, in addition to well-known mutations E484K, K417N, and N501Y,  mutations N439K, L452R,  S477N,  S477R, and  N501T are also the binding-strengthening mutations that have a high growth ratio recently with high frequency. Tracking the growth ratio tendency on a 10-day average for a long time enables us to detect the mutations that may strengthen the binding of S protein and ACE2, which will guide the development of vaccines  and antibody therapies. %The detailed information and a better view of plots can be found in the Supporting Information. 

Based on our early model of mutation impacts on antibodies \cite{chen2020prediction}, we found  that the E484K mutation may cause a dramatically disruptive effect on antibodies such as H11-D4, P2B-2F6, Fab 2-4, H11-H4, COVA2-39,  BD368-2, VH binder, S2M11, S2H13, CV07-270, P2C-1A3, P17, etc \cite{chen2020prediction}, which is consistent with the finding that E484K may affect neutralization by some polyclonal and monoclonal antibodies \cite{weisblum2020escape,resende2021spike}.  Mutation N501Y could weaken antibodies  B38, CC12.1, VH binder, S309 S2H12 S304, NAB, C1A-B12, C1A-F10, and STE90-C11 \cite{chen2020prediction}. 
Mutation N501Y could weaken antibodies B38, SR4, CC12.1, DB-604, S309 S2H12 S304, NABC1A-B12, etc.  Both E484 and N501 are coil residues on the RBD. Similarly, mutation K417N, which is a helix-residue of the RBD, could weaken antibodies B38, CB6, CV30, CC12.3, COVA2-04, BD-604, BD-236, NAB, P2C-1F11, C1A-B12, C1A-B3, C1A-F10, and C1A-C2, \cite{chen2020prediction}.
It is interesting to understand whether newly identified fast-growing mutations  N439K, L452R, and S477R are also disruptive to vaccines and antibodies. By checking the results reported early  \cite{chen2020prediction}, we note that mutation L452R may make antibodies H11-D4, P2B-2F6, SR4, MR17, MR17-K99Y, H11-H4, BD-368-2, CV07-270, and Fabs 298 52 ineffective. However, mutation N439K is not as disruptive as E484K, K417N,  N501Y, and N501T. It may weaken the binding of antibody SR4. S477N can slightly weaken antibodies DB23 and CV07-250. Finally, mutation S477R may even enhance the binding of most antibodies to the RBD.

\section{Methods}

\subsection{Data collection and pre-processing}
The first complete SARS-CoV-2 genome sequence was released on the GenBank ((Access number: NC\_045512.2)) on January 5, 2020, by Zhang's group at Fudan University \cite{wu2020new}. Since then, the rapid increment of the complete genome sequences is kept depositing to the GISAID database \cite{shu2017gisaid}. In this work, a total of 252,874 complete SARS-CoV-2 genome sequences with high coverage and exact submission date are downloaded from the GISAID database \cite{shu2017gisaid} (\url{https://www.gisaid.org/}) as of February 19, 2021. We take the NC\_045512.2 as the reference genome, and the multiple sequence alignment (MSA) will be applied by the Clustal Omega \cite{sievers2014clustal} with default parameters, which results in 252,874 SNP profiles.

\subsection{The growth rate of mutations}
Assume we have $N$ SNP profiles, which have a total of $M_n$ non-unique mutations and $M_u$ unique mutations ($M_u \le M_n$). Let $\Delta N_i$ be the number of the increment of a particular mutation during the $i$th 10-day period, and $N_i$ be the total number of a particular mutation.

Let the number of a particular mutation in the $j$th day of the $i$th 10-day period to be $N_{i}^j$, where $1\le i \le 10$. Let the $\Delta N_i = N_{i}^{10} -N_{i}^1$ be the number of the increment of a particular mutation during the $i$th 10-day period. Then the growth rate of a particular mutation in the $i$th 10-day period will be defined as
\begin{equation}
    R_j^i = 
    \begin{cases}
        0, \text{if }  \Delta N_i = 0 \text{ and } \sum_{k=1}^{i-1} \Delta N_k = 0, \\
        \frac{\Delta N_i}{(1+\sum_{k=1}^{i-1} \Delta N_k)}, \text{else}.
    \end{cases}
\label{eq:growth rate}
\end{equation}

Moreover, the natural logarithm growth rate of a particular mutation in the $i$th 10-day period will be defined as 
\begin{equation}
    LR_j^i = \text{log}(R_j^i + 1).
    \label{eq:log growth rate}
\end{equation}

\subsection{TopNetTree model for protein-protein interaction (PPI) binding free enrrgy changes upon mutation}
 Mutation-induced protein-protein binding free energy (BFE) changes are an important approach for understanding the impact of mutations on protein-protein interactions (PPIs) and viral infectivity \cite{li2020saambe}. A variety of advanced methods has been developed \cite{rodrigues2019mcsm,li2020saambe}. The topology-based network tree (TopNetTree) model \cite{wang2020topology,chen2020mutations} is applied to predict mutation-induced BFE changes of PPIs in this work. TopNetTree model was implemented by integrating the topological representation and network tree (NetTree) to predict the BFE changes ($\Delta\Delta G$) of PPIs following mutations  \cite{wang2020topology}.
The structural complexity of protein-protein complexes is simplified by algebraic topology \cite{carlsson2009topology,edelsbrunner2000topological,xia2014persistent} and is represented as the vital biological information in terms of topological invariants. NetTree integrates the advantages of convolutional neural networks (CNN) and gradient-boosting trees (GBT), such that CNN is treated as an intermediate model that converts vectorized element- and site-specific persistent homology features into a higher-level abstract feature, and GBT uses the upstream features and other biochemistry features for prediction. The performance test of tenfold cross-validation on the dataset (SKEMPI 2.0 \cite{jankauskaite2019skempi}) was carried out using gradient boosted regression tree (GBRTs). The errors with the SKEMPI2.0 dataset are 0.85 in terms of Pearson correlation coefficient ($R_p$) and 1.11 kcal/mol in terms of the root mean square error (RMSE) \cite{wang2020topology}. 

\subsubsection{Training set for TopNetTree model}
The TopNetTree model is trained by several important training sets. The most important dataset which provides the information for binding free energy changes upon mutations in the SKEMPI 2.0 dataset \cite{jankauskaite2019skempi}. The SKEMPI 2.0 is an updated version of the SKEMPI database, which contains new mutations and data from other three databases: AB-Bind \cite{sirin2016ab}, PROXiMATE \cite{jemimah2017proximate}, and dbMPIKT \cite{liu2018dbmpikt}. There are 7,085 elements including single- and multi-point mutations in SKEMPI 2.0. 4,169 variants in 319 different protein complexes are filtered as single-point mutations are used for TopNetTree model training. Moreover, SARS-CoV-2 related datasets are also included to improve the prediction accuracy after a label transformation. They are all deep mutation enrichment ratio data, mutational scanning data of ACE2 binding to the receptor-binding domain (RBD) of the S protein \cite{procko2020sequence}, mutational scanning data of RBD binding to ACE2 \cite{starr2020deep, linsky2020novo}, and mutational scanning data of RBD binding to CTC-445.2 and of CTC-445.2 binding to the RBD \cite{linsky2020novo}. Note the training datasets used in the validation in the main text does not include the test dataset, which the mutational data scanning data of RBD binding to CTC-445.2.

\subsubsection{Topology-based feature generation of PPIs}
Persistent homology, a branch of algebraic topology, is a powerful method for simplifying the structural complexity of macromolecules \cite{carlsson2009topology,edelsbrunner2000topological,xia2014persistent}. To construct topological data analysis on protein-protein interactions, we first preset the constructions for a PPI complex into various subsets. 
\begin{enumerate}%[ {(}1{)} ]
	\item $\mathcal{A}_\text{m}$: atoms of the mutation sites.
	\item $\mathcal{A}_\text{mn}(r)$: atoms in the neighborhood of the mutation site within a cut-off distance $r$.
	\item $\mathcal{A}_\text{Ab}(r)$: antibody atoms within $r$ of the binding site.
	\item $\mathcal{A}_\text{Ag}(r)$: antigen atoms within $r$ of the binding site.
	\item $\mathcal{A}_\text{ele}(\text{E})$: atoms in the system that has atoms of element type E. The distance matrix is specially designed such that it excludes the interactions between the atoms form the same set. For interactions between atoms $a_i$ and $a_j$ in set $\mathcal{A}$ and/or set $\mathcal{B}$, the modified distance is defined as
	\begin{equation}
	D_{\text{mod}}(a_i, a_j) =
	\begin{cases}
	\infty, \text{ if } a_i, a_j\in\mathcal{A}\text{, or }a_i, a_j\in \mathcal{B}, \\
	D_e(a_i, a_j), \text{ if } a_i\in\mathcal{A} \text{ and }a_j\in \mathcal{B},
	\end{cases}
	\label{eq:modified_equation}
	\end{equation}
	where $D_e(a_i, a_j)$ is the Euclidian distance between $a_i$ and $a_j$.
\end{enumerate}
In algebraic topology, molecular atoms of different can be constructed as points presented by $v_0$, $v_1$, $v_2$, $...$, $v_k$ as $k\!+\!1$ affinely independent points in simplicial complex. A simplicial complex is a finite collection of sets of points $K=\{\sigma_i\}$, and $\sigma_i$ are called linear combinations of these points in $\mathbb{R}^n$ ($n\ge k$). To construct a simplicial complex, the Vietoris-Rips (VR) complex and alpha complex, which are widely used for point clouds, are applied in this model \cite{edelsbrunner2000topological}. The boundary operator for a $k$-simplex would transfer a $k$-simplex to a $k-1$-simplex. Consequently, the algebraic construction to connect a sequence of complexes by boundary maps is called a chain complex
\[
\cdots \stackrel{\partial_{i+1}}\longrightarrow C_i(X) \stackrel{\partial_i}\longrightarrow C_{i-1}(X) \stackrel{\partial_{i-1}}\longrightarrow \cdots \stackrel{\partial_2} \longrightarrow C_{1}(X) \stackrel{\partial_{1}}\longrightarrow C_0(X) \stackrel{\partial_0} \longrightarrow 0
\]
and the $k$th homology group is the quotient group defined by
\begin{equation}
H_k = Z_k / B_k.
\end{equation}
Then the Betti numbers are defined by the ranks of $k$th homology group $H_k$ which counts $k$-dimensional invariants, especially, $\beta_0\!=\!{\rm rank}(H_0)$ reflects the number of connected components, $\beta_1\!=\!{\rm rank}(H_1)$ reflects the number of loops, and $\beta_2\!=\!{\rm rank}(H_2)$ reveals the number of voids or cavities. Together, the set of Betti numbers $\{\beta_0,\beta_1,\beta_2,\cdots \}$ indicates the intrinsic topological property of a system. 

Persistent homology is devised to track the multiscale topological information over different scales along a filtration \cite{edelsbrunner2000topological} and is significantly important for constructing feature vectors for the machine learning method. Features generated by binned barcode vectorization can reflect the strength of atom bonds, van der Waals interactions, and can be easily incorporated into a CNN, which captures and discriminates local patterns. Another method of vectorization is to get the statistics of bar lengths, birth values, and death values, such as sum, maximum, minimum, mean, and standard derivation. This method is applied to vectorize Betti-1 ($H_1$) and Betti-2 ($H_2$) barcodes obtained from alpha complex filtration based on the fact that higher-dimensional barcodes are sparser than $H_0$ barcodes.

\subsubsection{Machine learning models and training datasets}\label{Sec:ModelandTraining}
It is very challenging to predict binding affinity changes following mutation for PPIs due to the complex dataset and 3D structures. A hybrid machine learning algorithm that integrates a CNN and GBT is designed to overcome difficulties, such that partial topologically simplified descriptions are converted into concise features by the CNN module and a GBT module is trained on the whole feature set for a robust predictor with effective control of overfitting \cite{wang2020topology}.
The gradient boosting tree (GBT) method produces a prediction model as an ensemble method which is a class of machine learning algorithms. It builds a popular module for regression and classification problems from weak learners. By the assumption that the individual learners are likely to make different mistakes, the method using a summation of the weak learners to eliminate the overall error. Furthermore, a decision tree is added to the ensemble depending on the current prediction error on the training dataset. Therefore, this method (a topology-based GBT or TopGBT) is relatively robust against hyperparameter tuning and overfitting, especially for a moderate number of features.  The GBT is shown for its robustness against overfitting, good performance for moderately small data sizes, and model interpretability. The current work uses the package provided by scikit-learn (v 0.23.0) \cite{pedregosa2011scikit}. A supervised CNN model with the PPI $\Delta\Delta G$ as labels is trained for extracting high-level features from $H_0$ barcodes. Once the model is set up, the flatten layer neural outputs of CNN are feed into a GBT model to rank their importance. Based on the importance, an ordered subset of CNN-trained features is combined with features constructed from high-dimensional topological barcodes, $H_1$ and $H_2$ into the final GBT model.

\section{Conclusion}

Understanding the evolution trend of severe acute respiratory syndrome coronavirus 2 (SARS-CoV-2) and estimating its threats to the existing vaccines and antibody drugs are of paramount importance to the current battle against coronavirus disease 2019 (COVID-19). To this end, we carry out a unique analysis of mutations on the spike (S) protein receptor-binding domain (RBD). Our study is based on comprehensive 252,874 SARS-CoV-2 genome isolates recorded on the Mutation Tracker (\url{https://users.math.msu.edu/users/weig/SARS-CoV-2_Mutation_Tracker.html}). There are 5,420 unique single mutations and 650,852 non-unique mutations on the S protein gene. Therefore, an average genome sample has 2.6 mutations on the S protein but new samples have increasingly more mutations. In terms of the protein sequence, 535  missense mutations and 506 non-degenerate mutations occurred on the RBD. However, most of these RBD mutations have a relatively low frequency, leaving 95 significant mutations that have been detected more than 10 times in the database. We track fast-growing (FG) RBD mutations in 30 pandemic-devastated countries, including the UK, the US, Singapore, Spain, South Africa, Brazil, etc. To avoid random low-frequency mutations, we pursue this task by analyzing the 10-day growth rate of 95 significant RBD mutations.  We show that three fast-growing mutations N439K, L452R, S477N,  S477R, and N501T in addition to all known infectious variants containing N501Y, E484K, and K417N, deserve the world's attention. 

Additionally, we reveal that 92.6\% (88 out of 95) significant mutations on the RBD strengthen  the RBD binding  with the host angiotensin-converting enzyme 2 (ACE2), based on a cutting-edge topology-based neural network tree (TopNetTree) model trained on SARS-CoV-2 experimental datasets \cite{wang2020topology,chen2020prediction}. More specifically, we found that mutations N501Y, E484K, and K417N in the United Kingdom (UK), South Africa, or Brazil variants as well as mutations N439K, L452R, S477N,   S477R, and N501T are all associated with the enhancement of the BFE of the S protein and ACE2, confirming the earlier speculation. This result suggests that SARS-CoV-2 has evolved into more infectious strains due to the wide-spread transmission. 

Finally,  the early finding shows that more 70\% mutations would weaken the efficacy of known antibodies  \cite{chen2020prediction}. We report that rapidly growing mutations E484K, K417N, and L452R are more likely to disrupt existing vaccines and many antibody drugs, while mutations N501Y and N501T can also be disruptive, but mutations N439K, V367F, and S477R are not as disruptive as other rapidly growing ones. {
We have predicted vaccine escape mutations that are not only fast-growing but also can disrupt many existing vaccines. We have also identified vaccine weakening mutations as fast-growing RBD mutations that will weaken the binding between the S protein and many existing antibodies. A list of vaccine escape and vaccine weakening RBD mutations are predicted.} 
We unveil that regulated by host gene editing, viral proofreading,  random genetic drift, and natural selection,  the mutations on the S protein RBD tend to disrupt the existing antibodies and vaccines and increase the transmission and infectivity of SARS-CoV-2. 

\section*{Data and model availability}
The SARS-CoV-2 SNP data in the world is available at \href{https://users.math.msu.edu/users/weig/SARS-CoV-2_Mutation_Tracker.html}{Mutation Tracker}. The SARS-CoV-2 S protein RBD SNP data in 30 countries can be downloaded from the Supplementary Data. The TopNetTree model is available at \href{https://codeocean.com/capsule/2202829/tree/v1}{TopNetTree}. The related training datasets are described in Section \ref{Sec:ModelandTraining}. 

\section*{Supporting information}
The supporting information is available for \\
S1 BFE changes following 506 non-degenerate mutations on the S protein RBD.\\
S2 Supplementary Data. The Supplementary Data.zip contains two folders:
1. SNP Data: A total of 30 CSV files for the SARS-CoV-2 S protein RBD SNP data from 30 different
countries.
2. Fast Grow: A total of 30 HTML files for the log growth rates and log frequencies of specific SARS-CoV-2 S protein RBD mutations in 30 different countries.
\\
S3 Supplementary Figures. Figure S6 - Figure S25 plot the log growth ratio and log frequency of mutations on S protein RBD in the Australia, Austria, Belgium, Canada, Chile, China, France, Germany, Iceland, Ireland, Israel, Italy, Japan, Luxembourg, Norway, Portugal, Spain, Sweden, Switzerland, and the United Arab Emirates.

\section*{Acknowledgment}
This work was supported in part by NIH grant  GM126189, NSF grants DMS-2052983,  DMS-1761320, and IIS-1900473,  NASA grant 80NSSC21M0023,  Michigan Economic Development Corporation,  George Mason University award PD45722,  Bristol-Myers Squibb 65109, and Pfizer.
The authors thank The IBM TJ Watson Research Center, The COVID-19 High Performance Computing Consortium, NVIDIA, and MSU HPCC for computational assistance. RW thanks Dr. Changchuan Yin for useful discussion. 

%\bibliographystyle{abbrv}
%%% \bibliographystyle{custom}
%\bibliography{refs}
%% \end{multicols}

\end{document}